**Inflation Reduction Act impacts on the economics of clean hydrogen and synthetic liquid fuels**

Fangwei Cheng[a*], Hongxi Luo[a], Jesse D. Jenkins[a,b], and Eric D. Larson[a]

[a]Andlinger Center for Energy and the Environment, Princeton University, Princeton, NJ, 08544, USA

[b]Department of Mechanical and Aerospace Engineering, Princeton University, Princeton, NJ, 08544, USA

*Email: fangweic@princeton.edu

**Abstract:**

The Inflation Reduction Act (IRA) in the United States provides unprecedented incentives for deploying low-carbon hydrogen and liquid fuels, among other low greenhouse gas (GHG) emissions technologies. To better understand the prospective competitiveness of low-carbon or negative-carbon hydrogen and liquid fuels under the IRA in the early 2030s, we examine the impacts of IRA provisions on costs of producing hydrogen and synthetic liquid fuel made from natural gas, electricity, short-cycle biomass (agricultural residues), and corn-ethanol. With IRA credits (45V or 45Q), but excluding incentives provided by other national or state policies, hydrogen produced by electrolysis using carbon-free electricity (green $H_2$) and natural gas reforming with carbon capture and storage (CCS) (blue $H_2$) are cost-competitive with the carbon-intensive benchmark gray $H_2$ from steam methane reforming. Biomass-derived $H_2$ with or without CCS is not cost-completive under current IRA provisions. However, if IRA allowed biomass gasification with CCS to claim a 45V credit for carbon-neutral $H_2$ and a 45Q credit for negative biogenic-$CO_2$ emissions, this pathway would be less costly than gray $H_2$. The IRA credit for clean fuels (45Z), currently stipulated to end in 2027, would need to be extended, or similar policy support provided by other national or state policies, for clean synthetic liquid fuel to be cost-competitive with petroleum-derived liquid fuels. Levelized IRA subsidies per unit of $CO_2$ mitigated for all hydrogen and synthetic liquid fuel production pathways, except electricity-derived synthetic liquid fuel, range from 65 to 384 \$/t $CO_2$, which is within or below the range in U.S. federal government estimates of the Social Cost of Carbon (SCC) in the 2030 to 2040 timeframe.

**Synopsis statement:** Prospective cost-competitiveness of clean hydrogen and liquid fuels versus conventional fossil-fuel derived alternatives under the Inflation Reduction Act

**Keywords:** Inflation Reduction Act (IRA); Energy Policy; Hydrogen; Synthetic liquid fuels; Bioenergy with carbon capture and storage (BECCS); Electrofuels; Economics

**Introduction:**

In the United States, a commitment to cut greenhouse gas (GHG) emissions to half of the peak levels by 2030 and to net-zero by mid-century has become a major driver for new laws to accelerate the clean energy transition[1-3]. The Inflation Reduction Act (IRA) of 2022 provides unprecedented incentives for deploying low GHG emission technologies. The incentives include investment and production tax credits (ITCs and PTCs) designed to encourage businesses to invest in clean energy technologies.[*] The IRA tax credits and other mechanisms are projected to result in an estimated \$370 billion, and potentially more, of federal government outlays toward this goal over ten years (2023-2032)[4], including 45V credits for hydrogen produced with life cycle carbon intensity below 4 kg $CO_2$e/kg $H_2$, 45Q credits for $CO_2$ capture, utilization, and storage (CCUS), 40B and 45Z credits for low-carbon emission liquid fuels, and 48D and 45Y credits for carbon-free electricity generation, e.g., from nuclear, wind, or solar. The IRA also stipulates penalties for emissions of methane from specific types of facilities, primarily in the oil and gas sector[5, 6].

---

[*] "Clean" in this paper is short-hand for low-carbon or negative-carbon emissions. Other factors, such as criteria air pollutants (e.g., NOx, $SO_X$, and particular matter) are not considered in the analysis.

Several macro-energy system modeling studies have estimated that investments stimulated by the IRA would help to reduce U.S. GHG emissions to 33 – 40% below 2005 levels by 2030[7-10]. These studies estimate that GHG emission reductions under the new law come largely from the decarbonization of the power sector via wind and solar power generation[7, 8]. The Act significantly boosts investment in wind power and solar PV, almost doubling the annual capital investment from $181 billion under existing policy (i.e., no IRA) to $334 billion in 2030[7]. In addition to investment in the power sector, modeling results also show the IRA driving annual capital investments of $60 billion by 2035 in hydrogen and $30 billion in synthetic liquid fuel (SLF), including electrofuels and biomass-derived SLF[7].

Many modeling studies emphasize the crucial roles of hydrogen and SLF in clean energy transitions[11-13]. $H_2$ is currently an important feedstock used in ammonia production and oil refining[14]. In addition to existing demands, $H_2$ has the potential to play a critical role in decarbonizing steel production[12, 15], power generation[12, 16], transportation[17], and the petrochemical industry,[12, 18] among others. For example, steel produced by clean $H_2$-based direct reduction of iron - electric arc furnace processes result in 60% GHG emissions reduction as compared to conventional blast furnace-basic oxygen furnaces[15]. Biofuels and SLF can be drop-in alternatives to petroleum-derived liquid fuels (e.g., jet fuel, diesel, and gasoline). Since SLF can generally be utilized without requiring radical changes to fuel distribution and usage systems, they represent promising paths to displacing petroleum-derived liquid fuels for heavy-duty trucks, ships, and aircraft, thus reducing hard-to-abate emissions from the transportation sector[12].

Currently, natural gas reforming is the primary process for hydrogen production in the U.S., with life cycle GHG emissions of 10-12 kg $CO_2$e/kg $H_2$, depending on the conversion efficiency and level of upstream methane emissions[19]. Direct use of petroleum-derived liquid fuels (e.g., jet fuel, distillates, and gasoline) accounts for about 88% of U.S. transportation fuel usage, while biofuel (primarily corn-derived ethanol) accounts for 5%[20]. Without policy support, nascent clean $H_2$ and SLF production technologies today are not competitive with their carbon-intensive fossil-fuel counterparts. The IRA and other policy support is aimed at helping clean options gain early footholds in markets, scale-up commercial deployment, and achieve cost reductions via economies of scale, experience, and incremental innovation, as support policies have achieved for other technologies, e.g., wind generation[21].

In our previous study[22], we analyzed technologies for $H_2$ and Fischer-Tropsch liquid (FTL) production and determined their environmental and economic performance in the context of an assumed universally-applied carbon tax. Here, building on that work, we seek to provide a timely and nuanced analysis of specific provisions of the IRA affect the cost-competitiveness of low-carbon and negative carbon hydrogen and SLF production relative to carbon-intensive fossil fuel equivalents. The needed energy transition will be facilitated if clean $H_2$ and SLF are less costly to produce than fossil fuel counterparts. Other policies, including the federal Renewable Fuel Standard (RFS)[23] and California's Low Carbon Fuel Standard (LCFS)[24], have been promoted clean fuels production for transportation prior to passage of the IRA. Credit values under these programs have fluctuated significantly during the past 10 years, ranging from 0.05 – 3 $/RIN in the RFS program[25], and 20 to 210 $/t $CO_2$ in the LCFS program[26], as discussed in Section S2 of the Supplemental Information (SI).† Fuel producers can collect these incentives in combination with IRA credits, and we also assess potential outcomes of doing so.

In this article, we present plant-level techno-economic assessments to determine the levelized cost of fuel production (LCOF) for each of 15 different clean hydrogen or SLF production technological pathways (Table 1), taking into consideration provisions of the IRA (Supporting Information (SI), Table S1), RFS, and LCFS. We compare the LCOFs of clean $H_2$ and SLF with production costs of equivalent fossil benchmarks (gray $H_2$ and petroleum-derived liquid fuel). The timeframe of relevant IRA credits, their monetary values, and relevant statutory requirements are summarized in Table S1.

---

† Unless otherwise indicated all prices and costs in this paper are in 2022 dollars.

We seek to answer the following questions regarding clean hydrogen and clean liquid fuels:

(1) Can the IRA incentives make them competitive on a production-cost basis in the 2030 timeframe with conventional, carbon-intensive fossil-fuel based equivalents?

(2) Does the IRA appear to favor some production technology pathways over others?

(3) Are there shortcomings in the design of the IRA incentives for production of these, and if so, what modifications might help mitigate those?

Table 1. Summary of evaluated technologies for $H_2$ and SLF production and applicable policies.

| | Technology[1] | Feedstock | Relevant IRA credits[2] | Other policy[3] |
|---|---|---|---|---|
| **Hydrogen ($H_2$)** | | | | |
| P1 SMR | Steam methane reforming | Natural gas[4] | N/A | N/A |
| P2 SMR-CCS | Steam methane reforming w/CCS | Natural gas[4] | 45Q or 45V | N/A |
| P3 ATR-CCS | Autothermal reforming w/CCS | Natural gas[4] | 45Q or 45V | N/A |
| P4 Electrolysis | Electrolysis | Elec. (wind/solar)[5] | 45V | N/A |
| P5 BG-H2 | Gasifier + water-gas-shift | Biomass[6] | 45V | N/A |
| P6 BGCCS-H2 | Gasifier + water-gas-shift w/CCS | Biomass[6] | 45Q or 45V | N/A |
| **Synthetic Liquid Fuels (SLF)[7]** | | | | |
| *Synthesis from clean $H_2$ and carbon-neutral $CO_2$* | | | | |
| P7 SMR-SLF | SMR-CCS + reverse water-gas-shift-Fischer Tropsch synthesis (RWGS-FTS) | $H_2$ (P2), $CO_2$ DAC (direct air capture)[8] | 45Q or 45V for $H_2$, 45Q for $CO_2$ from DAC, 45Z | LCFS |
| P8 ATR-SLF | ATR-CCS + RWGS-FTS | $H_2$ (P3), $CO_2$ DAC[8] | 45Q or 45V for $H_2$, 45Q for DAC, 45Z | LCFS |
| P9 Elec-SLF | Electrolysis + RWGS-FTS | $H_2$ (P4), $CO_2$ DAC[8] | 45V for $H_2$, 45Q for DAC, 45Z | LCFS |
| P10 BG-SLF | BG-H2 + RWGS-FTS | $H_2$ (P5), bio-$CO_2$ | 45V for $H_2$, 45Z | LCFS |
| P11 BGCCS-SLF | BGCCS-H2 + RWGS-FTS | $H_2$ (P6), bio-$CO_2$ | 45Q or 45V for $H_2$, 45Z | LCFS |
| *Integrated biomass-gasification/FT synthesis* | | | | |
| P12 Int. Bio-SLF | Integrated biogasifier-FTS | Biomass | 45Z | RFS (D3),[10] LCFS |
| P13 Int. BioCCS-SLF | Integrated biogasifier-FTS w/ CCS | Biomass | 45Q or 45Z | RFS (D3),[10] LCFS |
| *Ethanol-to-Synthetic liquid fuel* | | | | |
| P14 EtOH-SLF | Corn-ethanol[9] to SLF | Corn-ethanol | N/A | RFS (D6),[10] LCFS |
| P15 EtOHCCS-SLF | Corn-ethanol[9] w/ CCS to SLF | Corn-ethanol w/ CCS | 45Q and 45Z | RFS (D5),[10] LCFS |

1. Pathways P1-P6 and P12-P13 represent production occurring at single facilities exploiting the single most remunerative IRA incentive among those for which they are eligible. The IRA disallows any one facility from claiming more than one of 45V, 45Q, and 45Z credits. Pathways P7-P9 are assumed to consist of 3 facilities each, which enables these pathways to incorporate 3 separate IRA credits: 45V or 45Q for $H_2$ production, 45Q for direct air capture, and 45Z for SLF production. Pathways P10-P11 each include two facilities, one producing $H_2$ and $CO_2$ from biomass and garnering 45V or 45Q credits and the other combining $H_2$ and $CO_2$ from the first facility to synthesize liquid fuels and collect 45Z credits. Similarly, pathway P15 includes capture of fermentation $CO_2$ at a standalone ethanol facility claiming 45Q for CCS, and a separate ethanol to SLF facility claiming 45Z.
2. In addition to these listed credits, our analysis embeds IRA 45Y credits for clean electricity production into the assumed electricity price for pathway P4 and the IRA's stipulated penalties for methane emissions into the assumed natural gas price for pathways P1-P3. (See Table S1 for descriptions of IRA credits.) All pathways are assumed to satisfy prevailing wage and apprenticeship requirements and thus qualify for the maximum level of credit within a given type of IRA credit.

3. Other policies considered in this study are the LCFS and RFS, but only for the SLF pathways. Hydrogen used in transport may be eligible for LCFS credits [27], but we do not consider this option (see Limitations section of the paper). The P7-P15 pathways produce drop-in transportation fuels, and we assume all are eligible for LCFS credits. Integrated biomass-gasifier-FTS (P12 and P13) are eligible for RFS since they produce SLF directly from lignocellulosic biomass.

4. We assume upstream natural gas emissions of 1.7% (0.29 g of methane per MJ of delivered natural gas)[28]. The IRA methane emissions penalty is $1500/tonne starting in 2025 (Table S1). We reflect this in the natural gas price assumed in our analysis ($ 4.3/GJ, see Table S3). Upstream life cycle upstream GHG emissions of natural gas are 13.8 kg $CO_2$/$GJ_{HHV}$ (Table S4) and it's carbon content is 50 kg $CO_2$/$GJ_{HHV}$ (Table S5).

5. The electrolysis pathway is assumed to use wind- and/or solar-derived electricity as input. The 45Y clean electricity PTC is incorporated into the electricity price assumed in our analysis. The input electricity price after applying the 45Y subsidy is $ 21.5/MWh (see Table S3) and the capacity factor for electrolysis is 0.55 [13, 22] as stipulated in modeling of Larson, et al [13].

6. The biomass used here is assumed to be agricultural residues, and we treat its biogenic $CO_2$ to be carbon-neutral, consistent with its treatment in the billion-ton study.[29] Upstream (non-biogenic) emissions of 5.08 kg $CO_2$/GJ (Table S4) are assumed from harvesting, transport, and pre-processing. The delivered cost of the biomass is 6.1 $/GJ (Table S3), and its carbon content is 87.9 kg $CO_2$/$GJ_{HHV}$ (Table S5).

7. A variety of SLF compositions are possible, depending on process design. Here we assume the composition to be 82% synthetic paraffinic kerosene (SPK) and 18% naphtha (lower heating value basis), which is composition determined in a prior detailed process modeling study involving one of this paper's authors[30].

8. For P7-P9, the requisite carbon-neutral $CO_2$ input is assumed to be from direct air capture (DAC) powered by renewable energy. The IRA's 45Q credit (130 $/t $CO_2$ for 12 years) is applied to reduce the cost of $CO_2$ from DAC, giving $CO_2$ costs of 164 $/t $CO_2$ (see note *e* in Table S3).

9. The corn-ethanol is assumed to be 90% from dry mills and 10% from wet mills (the US industry average from GREET[31]), with a delivered cost of $1.83/gal (Table S3). For corn-ethanol with CCS, $CO_2$ from fermentation is compressed transported, and stored at an added cost of 45$/t $CO_2$ (Table S3).

10. Pathways P12 and P13 can claim RFS cellulosic biofuel (D3) credits. Pathways P14 and P15 are eligible for RFS renewable fuel (D6) and advanced biofuel (D5) credits, respectively, based on the GHG emissions reductions they provide.

**Method:**

The timeframe for relevant IRA credits, their values for different carbon intensity products, and other relevant statutory requirements are summarized in Table S1. Building on details provided in our previous study[22], the performances and costs of technologies prospectively commercially deployed in the early 2030s are summarized in Table 2. Table S2 provides additional details. For pathways P1 through P11 we used performance and cost data and methods from our previous study (Table 2), but with adjustments to LCOF calculations to incorporate the impact of provisions of the IRA (Table S1) and of the LCFS and RFS (SI section 3). Pathways P12 to P15 were not included in our previous study. We developed performance and cost metrics for these consistent with those in our prior study, as described in notes *g* and *h* of Table S2. All facilities are assumed to start operating by 2032 and have book lives of 15 years. We have assumed plants start up in the early 2030s because 1) IRA credits accrue only to a producer who commences construction by the end of 2032 and 2) the time available between today and 2032 seems sufficient for commercial projects to be launched by then.

We use a derating factor (DF) to account for the impacts of different policy durations (e.g., 10 years for 45V and 12 years for 45Q) over evaluated 15-year project lifetimes. A DF of 1 means that the provision is available during the entire project lifetime. DF is determined by Eqn (1),

$$DF = \frac{\sum_1^m (1/(1+r))^m}{\sum_1^n (1/(1+r))^n} \quad (1)$$

where *r* is the real weighted-average cost of capital (WACC), *n* is the book life of the facility (15 years), and *m* is the duration of the IRA provision (Table S1). A WACC value of 0.1 is used, as used in modeling $H_2$ and SLF production technologies by Jones and Haley in Annex A.2 of Net Zero America report[13].

The LCOFs for hydrogen pathways are determined by Eqn (2),

$$LCOF_{p,H2}(\$/kg) = \frac{CAPEX_p * CRF + FOM_p + VOM_p + F_p + Seq_p}{M_{p,H2}} \quad (2)$$

$$- \frac{Max(M_{ccs,p} * 45Q_p * DF_{45Q}, M_{H2,p} * 45V_p * DF_{45V})}{M_{p,H2}}$$

where subscript $p$ refers to a specific $H_2$ generation pathway (P1-P6), and *CAPEX* is the overnight capital cost ($) of the facility, *CRF* is the annual capital recovery factor (0.131), *FOM* is the annual fixed operating and maintenance cost ($), *VOM* is the annual non-fuel variable cost ($), *F* is the annual cost of feedstock (natural gas, electricity, or biomass) ($), *Seq* is the annual $CO_2$ transport and storage cost ($), $M_{H2}$ is the annual hydrogen production (kg), $M_{ccs}$ is the annual $CO_2$ sequestered (tonne), *45Q* is the tax credit for CCUS ($/t$CO_2$ based on Table S1), *DF* is the derating factor (Eqn. 1), and *45V* is the tax credit for clean hydrogen production ($/kg, as in Table S1). Hydrogen product leaving the facility gate is at 30 bar. For all hydrogen production pathways other than electrolysis (P4), the final hydrogen product is delivered from pressure-swing adsorption (PSA) units, which typically operate at a pressure around 30 bar[32]. For P4, the hydrogen product is delivered directly from PEM electrolyzers, which are capable today of delivering $H_2$ at 30 bar[33]. The capacity factor for $H_2$ routes is assumed to be 0.85, except for electrolysis, which is assumed to be 0.55 (see note 5 of Table 1).

The SLF output of P7 to P15 is assumed to 82% synthetic paraffinic kerosene (SPK) and 18% naphtha on a lower heating value basis. (See Table 1, note 7.) The 45Z credit is $1.75/gal and $1/gal for sustainable aviation fuel (SAF) and other types of clean fuels (e.g., synthetic diesel), respectively. Therefore, the 45Z tax credit for 1 gallon of SLF in our analysis is 1.75*0.82 + 1*0.18 = $1.62/gal, before applying any DF value. 1 gal of SLF has a lower heating value (LHV) of 0.126 GJ. LCOFs of liquid fuel pathways are determined by Eqn (3),

$$LCOF_p\left(\frac{\$}{gal}\right) = \frac{CAPEX_p * CRF + FOM_p + VOM_p + F_p + Seq_p}{V_{p,SLF}} + \frac{M_{dac,p} * (P_{dac} - 45Q_{dac} * DF_{45Q})}{V_{p,SLF}} \quad (3)$$

$$- \frac{M_{co2\,ethanol,p} * (45Q_{ccs} * DF_{45Q})}{V_{p,SLF}} - (EF_p * DF_{45Z} * 45Z) - RFS_p - LCFS_p$$

where subscript $p$ refers to a specific SLF pathway (P7-P15), and *CAPEX, CRF, FOM, VOM, Seq* are analogously defined as for Eqn. 2, *F* is the annual cost of feedstock ($, for biomass, ethanol, or $H_2$, where $H_2$ cost is determined by Eqn 2), $V_{SLF}$ is the annual SLF production (gallons), $M_{dac}$ is the annual external $CO_2$ from DAC (tonnes) used for SLF production, $P_{dac}$ is the price of the $CO_2$ from DAC (i.e., 280 $/t $CO_2$, note *e* of Table S3), $45Q_{dac}$ is the tax credit for $CO_2$ utilization from a DAC facility (i.e., 130 $/t $CO_2$, Table S1), $M_{co2,\,ethanol}$ (tonnes) is the annual $CO_2$ captured from a corn-ethanol facility, $45Q_{ccs}$ is the tax credit for $CO_2$ captured and sequestrated from corn-ethanol CCS facilities (i.e., 85 $/t $CO_2$, Table S1), *45Z* is the full tax credit for SLF production (i.e., 1.62 $/gal), *EF* is the emissions factor for calculating the 45Z credit (Table S1), *DF* is the derating factor (Eqn. 1), *RFS* is the RIN credit SLF can receive under the RFS program ($/gal, Eqn. S1), and *LCFS* is the carbon credit under California's LCFS program ($/gal, Eqn. S2). Carbon footprints for assessing 45Z credit values are statutorily determined using the GREET model, which uses LHV values by default[31]. Therefore, we convert the life cycle GHG emissions given in Table S4 to a LHV basis (kg $CO_2$/MMBTU$_{LHV}$) and use these to determine EF values. The capacity factor for all the SLF pathways is assumed to be 0.85.

The CAPEX, *FOM*, *VOM*, and energy conversion efficiency associated with all the pathways are given in Table 2. See Table S2 for supporting details, including data sources, assumptions, and calculations, Table S3 and Table S4 for input energy and materials prices and life cycle GHG emissions.

Table 2. Summary of evaluated hydrogen and SLF technologies. Parameter values here are given per unit of output capacity or output energy. Energy values are on a higher heating value (HHV) basis. IFI is input energy divided by output energy (HHV).

Co-product is electricity for all the cases, and positive values are inputs; negative values are outputs. See SI Section 3 (Table S2) for supporting details.

|  | Technology | Input Feedstock Intensity (IFI) ($GJ_{in}/GJ_{out}$) | Co-product (GJ/GJ) | $CO_2$ capture rate (%) | CAPEX ($/kW capacity) | Fixed O&M ($/kW-year) | Variable O&M ($/GJ) |
|---|---|---|---|---|---|---|---|
| $H_2$ | SMR | 1.23 (NG) | 0 | 0 | 543 | 19 | 0.36 |
|  | SMR-CCS | 1.26 (NG) | 0.024 | 91.2 | 943 | 28 | 0.00 |
|  | ATR-CCS | 1.2 (NG) | 0.048 | 94.1 | 825 | 25 | 0.00 |
|  | Electrolysis | 1.28 (electricity) | 0 | 0 | 1407 | 16 | 0.00 |
|  | BG-H2 | 1.78 (biomass) | - 0.079 | 0 | 2482 | 44 | 3.89 |
|  | BGCCS-H2 | 1.78 (biomass) | - 0.023 | 87 | 2587 | 46 | 3.89 |
| SLF | RWGS+FTS | 1.47 ($H_2$) | 0 | 0 | 1004 | 27 | 0.73 |
|  | Integrated BioFTS | 2.02 (biomass) | 0 | 0 | 3826 | 77 | 5.15 |
|  | Integrated BioFTS-CCS | 2.02 (biomass) | 0 | 87 | 3944 | 79 | 5.15 |
|  | Ethanol-to-Jet | 1.09 (ethanol) | 0 | 0 | 258 | 26 | 2.15 |
|  | Ethanol with CCS-to-Jet | 1.09 (ethanol) | 0 | 0 | 258 | 26 | 2.15 |

Finally, Levelized Subsidies for $CO_2$ Mitigation (LSCM) implied in IRA credits for clean $H_2$ or SLF are determined by Eqn 4.

$$Levelized\ subsidies\ of\ CO_2\ mitigation_p \left(\frac{\$}{CO_2e}\right) = \frac{-NG\ Fee_p + 45V_p + 45Q_p + 45Z_p + 45Y_p}{CI_{fossil} - CI_p} \quad (4)$$

For each pathway (*p*), the credits (45*V*, 45*Q*, 45*Z*, and 45*Y*) and the methane emissions penalty (*NG Fee*) are converted into $/kg $H_2$ or $/gal SLF of final product. $CI_p$ is the lifecycle carbon intensity of the pathway expressed in kg $CO_2$e/kg$H_2$ or kg $CO_2$e/gallon liquid fuel. $CI_{fossil}$ represents the carbon intensity of the fossil benchmark. For $H_2$, this is the CI of pathway P1 (11 kg $CO_2$e/kg $H_2$) and for SLF $CI_{fossil}$ is 10.7 kg $CO_2$e/gal (representative of conventional jet fuel).

**Results:**

**Hydrogen Production Pathways.** Figure 1A displays hydrogen LCOFs, and corresponding life cycle GHG emissions.

Upstream natural gas supply chain emissions and uncaptured $CO_2$ result in life cycle GHG emissions for blue hydrogen pathways, SMR-CCS (P2) and ATR-CCS (P3), of 3.3 and 2.9 kg $CO_2$e/kg $H_2$, respectively. Hence, these two pathways only qualify for the lowest level 45V credit, i.e., $0.6/kg $H_2$ (Table S1), and benefit more from taking the 45Q credit. The resulting LCOF of blue hydrogen via P2 SMR-CCS ($1.24/kg $H_2$) and P3 ATR-CCS ($1.16/kg $H_2$) are slightly lower than for the benchmark gray hydrogen (P1) ($1.29/kg $H_2$). Hydrogen from electrolysis using solar- or wind-derived electricity has zero carbon emissions and so receives the maximum 45V credit ($3/kg $H_2$), which reduces its LCOF to $0.31/kg $H_2$, well below the cost of gray hydrogen. Hydrogen via P5 BGH2 has life cycle emissions of 1.3 kg $CO_2$e/kg $H_2$ due to upstream GHG emissions associated with biomass production, processing, and transportation. Thus, the 45V credit available in this case is $1.00/kg $H_2$, and the LCOF is uncompetitive at $3.09/kg $H_2$. Finally, hydrogen via the P6 BGCCSH2 pathway has negative carbon emissions because biogenic $CO_2$ is captured and stored. Unlike the case with SMR-CCS and ATR-CCS, the $3/kg $H_2$ credit from 45V has a larger effect on LCOF than the $85/t $CO_2$ credit from 45Q. Thus, we apply 45V credits and obtain a LCOF of $2.22/kg $H_2$.

These results demonstrate that the IRA is likely to drive commercial-scale deployment of green and blue $H_2$. Hydrogen produced via BGCCSH2 ($2.22/kg $H_2$) is less economically attractive than green or blue hydrogen and costs significantly more than gray hydrogen. The 45V hydrogen credit is a tiered credit system but does not provide additional incentive for the negative-emissions that characterize BGCCSH2. Thus, under the IRA, the economics of BGCCSH2 could be improved by either limiting the

size of CCS equipment to capture and sequester just enough $CO_2$ to claim the full 45V credit or by capturing all $CO_2$ and selling the excess (about 95% of the total captured) to other facilities for revenue if there is a market for biogenic $CO_2$. If pursuing the latter strategy, the revenue from selling biogenic $CO_2$ for a price of about $45/t would bring the LCOF for BGCCSH2 to the same level as for gray $H_2$.

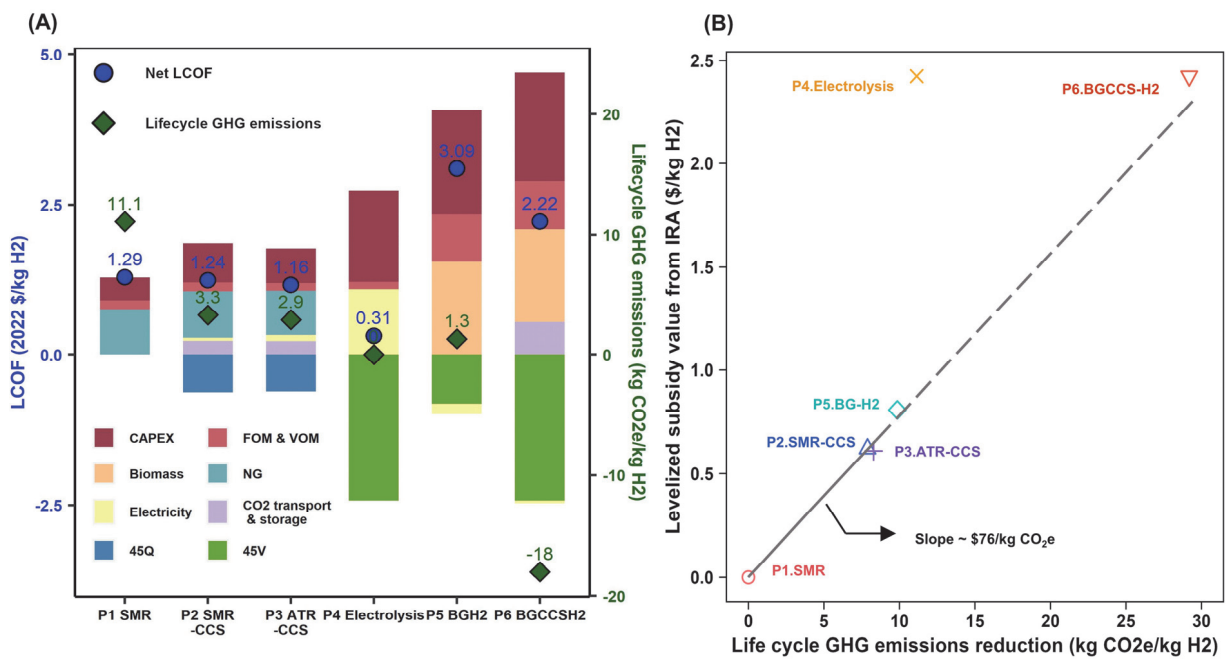

Figure 1. **(A)** Levelized cost of fuel production (LCOF), where positive values are costs and negative values are revenues from co-product sales or IRA credits, and life cycle GHG emissions for benchmark (P1.SMR) and five lower emissions hydrogen production pathways. Methane emission fees are embedded in our assumed natural gas price for P1- P3, and 45Y clean electricity credits are embedded in our assumed electricity price for P4 (see Table S3). The life cycle GHG emissions for P1-P6 (assuming $GWP_{100}$ for non-$CO_2$ GHGs) are adopted from our previous analysis [22], and 142 $GJ_{HHV}$/kg $H_2$ is used to convert from kg $CO_2e/GJ_{HHV}$ in[22] to kg $CO_2e$/kg $H_2$. IRA credits shown here are expressed as derated values to account for project economic lifetimes longer than the IRA-stipulated durations over which incentives are available, e.g., the $1/kg $H_2$ and $3/kg $H_2$ 45V credits are derated to $0.8/kg $H_2$ and $2.4/kg (See discussion around Eqn. 1 in Method.) **(B)** Levelized IRA subsidy values as a function of the life cycle GHG emissions reductions for clean hydrogen (P2-P6) versus the benchmark (P1).

For each of the hydrogen pathways, Figure 1B plots the levelized IRA 45V or 45Q incentive, expressed per kg of $H_2$ produced, against the pathway's lifecycle GHG emissions reduction relative to emissions from the benchmark pathway (P1). One observes that the levelized IRA subsidy for each pathway is not uniformly proportional to its emissions reduction. In particular, the electrolysis pathway (P4) departs markedly from proportionality. One likely explanation for this discrepancy is that IRA credits have two objectives: emissions reductions from hydrogen production in the near-term and technological advancements to reduce clean hydrogen costs in the long term. As society seeks to reach net-zero emissions by mid-century, it seems unlikely that fossil fuel-based blue hydrogen pathways (P2 and P3) will play major roles in the long run. We thus conclude that IRA incentives for the P2 and P3 pathways primarily incentivize near-term emissions reductions and not technology development, a conclusion further buttressed by the fact that the reforming technologies at the core of the P2 and P3 pathways are commercially mature today, and there are at least two commercial-scale reforming projects operating with $CO_2$ capture and storage and several additional projects have been announced[34, 35]. In contrast, electrolysis (P4) and biomass gasification (P5 and P6) technologies are not yet commercially mature and have potential for significant cost-reduction through further technology advances and learning. P4 and P6 also have zero or negative lifecycle GHG intensities and can thus play a key long-term role in net-zero emissions economies. One might logically expect IRA incentives for these technologies to thus include an emissions reduction component comparable to that for P2 or P3 plus an additional "technology policy" incentive to promote near-term technology advancement/cost reduction.

The slope of a best-fit line drawn in Figure 1B through the methane reforming pathways (P1 – P3) is $76/t $CO_2$e reduced, which could be interpreted as the amount of IRA subsidy provided strictly for reducing emissions. Extrapolating this line to the larger reductions in $CO_2$e emissions characterizing P4, P5, and P6 highlights the fact that the IRA incentive for green hydrogen (P4), $217/t$CO_2$e reduced, is nearly triple the level of incentive associated with emissions reductions alone. The additional IRA incentive aimed at spurring technological advancement/cost reduction in this case is $141/t$CO_2$e reduced. Meanwhile, the total IRA incentives for $H_2$ for biomass gasification (P5) and BGCCSH2 (P6) include negligible additional bonuses. It is difficult to know the full reasons the IRA authors included a large bonus incentive for electrolytic hydrogen and nearly no bonus for biomass gasification-derived hydrogen. A partial explanation may be that they considered electrolysis to be an inherently modular technology, whereas biomass gasification and CCS are not. Modularity is regarded as among the most important factors for enabling rapid cost learning[36], as evidenced by dramatic cost reductions and rapid market share gains of wind turbines and solar PV modules for electricity generation. Hence, a sizeable IRA bonus incentive for electrolysis is well aligned with the ambitious goals of rapid societal electrification coupled with deep decarbonization of the electricity sector. On the other hand, by providing only a negligible bonus incentive for biomass gasification w/CCS (P6), the pathway with the highest carbon mitigation potential, the IRA misses an opportunity to ensure early market deployment and scale-up of a technology that might also contribute significantly to rapid decarbonization and net-zero economy-wide-emission goals. If the IRA were to allow a biomass-hydrogen producer to earn both 45V credits for carbon-neutral hydrogen plus 45Q credits for negative biogenic $CO_2$ emissions, P6 hydrogen would be cost-competitive with gray (and blue) hydrogen with an LCOF of 0.84 $/kg.

**Synthetic liquid fuel pathways.** The SLF production pathways analyzed in this work (P7-P15) have three basic configurations. P7 through P11 use a RWGS-FTS process taking clean hydrogen and carbon-neutral $CO_2$ derived from different processes (Table 1) as inputs, with hydrogen costs corresponding to the net LCOFs in Figure 1A. For P7-P9, the input $CO_2$ is assumed to be from DAC plants, and the IRA's credit for DAC under 45Q is applied to the input $CO_2$ price, giving $164/$CO_2$ (Table S3). Pathways P10 and P11 synthesize fuels using $H_2$ from P5 and P6, respectively, and carbon-neutral biogenic $CO_2$ recovered from the P5 and P6 facilities. The SLF costs for P10 and P11 incorporate 45V credits associated with P5 and P6 facilities, which are assumed to be adjacent to, but separate from, the SLF production facility. $CO_2$ separation is intrinsic to making hydrogen from biomass, and so it is assumed that the SLF producer incurs no extra cost for the $CO_2$, beyond what is reflected in the cost of the delivered hydrogen.

P12 and P13 pathways use biomass as input into integrated gasifier-FTS facilities, of which several configurations are under commercial development today [37]. Note that pathways P10 and P11 represent another approach to using biomass to make SLF. The design of P10 and P11 is motivated strictly by the fact that the IRA disallows any one facility from claiming more than one credit: input $H_2$ from a dedicated biomass-to-hydrogen facility and $CO_2$ from the same facility (from the separated $CO_2$ stream) are combined and converted to SLF via FTS at a second facility. Each facility involved may claim an IRA credit, which effectively stacks two credits, whereas the biomass-to-SLF integrated single-facility configuration of P12 or P13 is eligible under the statute for, at most, one IRA credit.

P14 and P15 pathways are based on upgrading ethanol into SLF via dehydration of ethanol, oligomerization of the resulting ethylene into long-chain hydrocarbons, hydrogenation of the latter, and finally fractionation into desired fuels (e.g., naphtha and SPK)[38]. It is assumed that the ethanol used for P14 and P15 is produced at standalone facilities.

Unlike our analysis for hydrogen (Figure 1), we include no direct liquid fuel subsidies in our LCOF calculations for SLFs for two reasons. First, the IRA's 45Z credit that would apply to SLF is stipulated to expire in 2027, and unlike 45V or 45Q, the credit does not offer a multi-year payment stream for projects eligible for the credit at the time they commence construction. Instead, payments from 45Z are made only for eligible fuels produced in the years the credit is in effect. Unless the credit is renewed and extended it will therefore not be available to any facilities operating in the 2030s. Second, the federal RFS and California LCFS have been important driving factors to promote the deployment of low-carbon fuels in the transportation sector, but both credits are subject to high uncertainty and volatility (Figure S1). Hence, we have chosen to exclude all three of these policies from our baseline scenario analysis and instead to explore the impact of these as a sensitivity later in this paper.

Figure 2 presents LCOF and life cycle GHG emissions for all the SLF pathways. Using electrolytic $H_2$ and $CO_2$ from DAC to synthesize liquid fuels (P9, or "electro-fuels") results in the lowest LCOF ($2.9/gal). While higher than the median value of recent fossil jet fuel prices ($2.2/gal), this estimated LCOF is within the recent historical range in jet fuel prices and may therefore prove competitive without any explicit liquid fuel subsidy, especially for customers with voluntary advanced market commitments to purchase sustainable aviation fuel [39]. Fossil hydrogen based SLF pathways have much higher LCOFs at $4.3/gal (P7) and $4.2/gal (P8).

With bio-derived $H_2$ and $CO_2$ as feedstocks (P10 and P11) or with direct biomass feedstock (P12 and P13), LCOFs for SLF are generally higher, ranging from $4.2/gal to $6/gal. Comparing the two biomass-to-SLF approaches noted above, we find that SLF via P12 (integrated-gasification-FTS) and P13 (integrated-gasification-FTS with CCS) have higher LCOF than their corresponding 2-facilities pathways, P10 and P11, respectively. The single-facility configurations have lower CAPEX and higher energy efficiency than the 2-facilities configurations (see note *g* of Table S2), which means that the LCOF would be lower for P12 vs. P10 and for P13 vs. P11 in the absence of any subsidies. However, given the IRA's prohibition against a single facility earning more than one tax credit, it is economically preferable to pursue the more capital-intensive and less efficient process of producing hydrogen and $CO_2$ at one facility and producing SLF from the $H_2$ and $CO_2$ at a different facility in order to claim multiple IRA credits. This is concerning, considering that sustainable biomass feedstock is a limited resource. This finding is explored further in Section 5 of the SI.

As for GHG emissions for P10 and P12, where CCS is not part of the pathway, total GHG emissions are due only to upstream biomass production, transportation, and processing. Since the biomass-to-SLF conversion efficiency for P12 is higher than for the P10, less biomass input is required with P12 and the GHG emissions are correspondingly lower. When CCS is part of the process, as in P11 and P13, biogenic $CO_2$ is captured and sequestered, resulting in net negative lifecycle GHG emissions. In this case, the process that converts biomass to SLF with lower efficiency (P11) results in more byproduct $CO_2$ available for capture (and less converted to SLF) and thus has more-negative lifecycle GHG emissions than the more efficient pathway, P13.

Using ethanol to produce SLF (P14 and P15) results in moderate LCOFs ($3.8/gal and $3.6/gal) that straddle the historically highest price of jet fuel seen in the past decade. These LCOFs are higher than the LCOF of electro-fuels (P9) but are lower than that of other SLF pathways evaluated in this study. The life cycle GHG emissions of P14 are the highest among the SLF pathways, which is mostly due to emissions associated with corn-ethanol production, including emissions from corn farming and heat and electricity used for refining[40]. Capturing the fermentation $CO_2$ (pathway P15) reduces lifecycle emissions by nearly two-thirds, bringing them down nearly to the level of emission for the biomass-SLF facilities without CCS (P10 and P12).

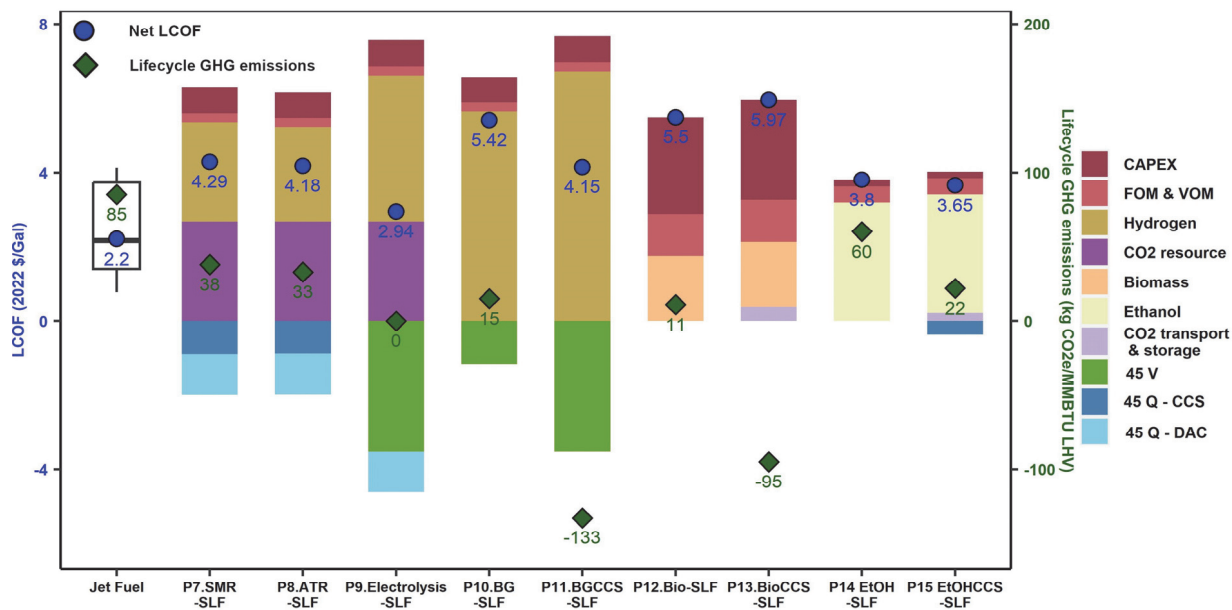

Figure 2. LCOF and life cycle GHG emissions (GWP$_{100}$) for SLF production pathways. The SLF product is assumed to consist of 82% synthetic paraffinic kerosene (SPK) and 18% naphtha on an LHV energy basis. (See Table 1, note 7.) The life cycle GHG emissions for P7-P11 are adopted from our previous analysis[22], and converted to kg CO$_2$e/MMBtu$_{LHV}$ shown here assuming a HHV/LHV ratio of 1.05. The life cycle GHG emissions of P12-P15 are determined using assumptions and methodology consistent with [22]. The left-most bar represents the range in annual-average wholesale petroleum jet fuel prices seen in the US from 2012 to 2022 [41]. The median value was $2.2/gallon. For P7-P11, the 45V and 45Q-CCS credits shown here are those that are incorporated in the LCOF of H$_2$ (Figure 1A), but expressed on a per-gallon of SLF basis, and the hydrogen cost plotted here is gross H$_2$ cost (i.e., with IRA credits excluded).

**Sensitivity analysis: low-carbon fuel subsidies.** For reasons noted earlier, our baseline SLF analysis excluded any consideration of 45Z, LCFS, or RFS credits. Here we explore how these subsidies might impact LCOF for SLF pathways.

To assess the impacts of extending 45Z into the 2030s, we determined the LCOF of each SLF pathway as a function of the 45Z duration, ranging from 0 to 15 years. Not surprisingly, longer durations yield lower LCOFs (Figure 3A). As noted earlier, the electrolysis-based pathway (P9), and the EtOHCCS-SLF pathway (P15), even without any 45Z credit, have LCOFs that falls within the recent historical price range for jet fuel. If the 45Z duration is at least two years, then bio-derived SLF (with H$_2$ and CO$_2$ from separate facilities, P11) also becomes competitive with historical jet fuel prices, while a 45Z term of five years or more would be needed to make bio-SLF at integrated facilities (P13) viable. The P11 pathway captures both 45V and 45Z credits, while P13 captures only a 45Z credit, which explains the lower LCOF for P11 than P13 at any given 45Z duration. Both P11 and P13 LCOFs fall rapidly with increasing durations of 45Z, because their negative lifecycle emissions (Figure 2) lead to very high 45Z credit values (Figure S2 and Eqn. S3). The P11 LCOF falls more rapidly than for P13, even though both pathways start with biomass and end with SLF produced. Because P11 has a lower overall energy conversion efficiency, the pathway captures more CO$_2$ per unit of SLF produced and so receives a larger 45Z credit than the more efficient P13 pathway. No other SLF pathways become economically viable with 45Z durations less than 15 years.

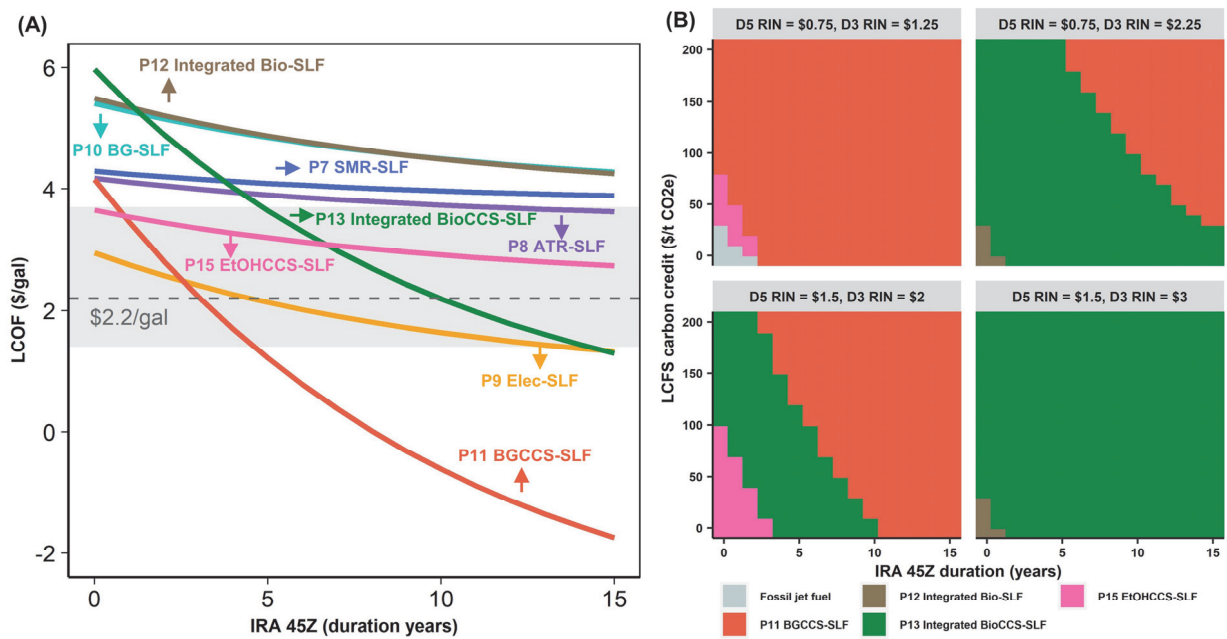

Figure 3. **(A)** LCOF for SLF production pathways as a function of 45Z duration. The shaded region represents the range in monthly average wholesale jet fuel prices in the U.S. from 2012 to 2022 [41]: the lower bound is at the 10th percentile of the range ($1.4/gal), and the upper bound is at the 90th percentile ($3.72/gal). P14 is not shown on the graph because it is ineligible to receive 45Z and thus is insensitive to the duration of 45Z. **(B)** SLF production pathways that would be competitive with fossil jet fuel at 2.2 $/gal under different 45Z durations, LCFS carbon credits, and RIN prices. All pathways (P7-P15) were considered in the analysis, but only P11, P12, P13, and P15 are competitive in the variable space examined here. P12 - P15 pathways are eligible for RIN credits (at 1.64 RIN credits per gallon). P14 and P15 are assumed to claim D6 and D5 RINs, respectively. Here we assume D6 and D5 RIN prices are the same. P12 and P13 are assigned D3 RINs, and the D3 price is the D5 price plus $0.5 and $1.5, which is the 25th and 75th percentile of the price difference between D3 RINs and D5 RINs in the past ten years. See Section 2 of the SI for discussion of RIN categories and prices.

In addition to 45Z, clean SLF may benefit from additional existing policies such as the RFS and LCFS, if these do not expire in the 2030s. Here we assume all pathways claim LCFS credits based on their carbon intensity (see Eqn. S2). The integrated biomass-SLF facilities (P12 and P13), and ethanol-SLF facilities may additionally claim Renewable Identification Number (RIN) credits under the category of cellulosic biofuel D3 (for P12 and P13), advanced biofuel D5 (for P15), and renewable biofuel D6 (for P14), as discussed in the Section 2 of the SI. Due to the uncertainty and volatility of RIN and LCFS credits (Figure S1), we examine LCOF values over a range of LCFS credit levels, RIN prices, and 45Z durations. Within this variable space, Figure 3B shows which pathways provide an LCOF competitive with a fossil jet fuel priced at $2.2/gal, the median historical value during 2012-2022.

With the D5 RIN at $0.75 and the D3 RIN at $1.25 (upper left panel in 3B), fossil jet fuel is the least costly option when the LCFS credit is below $25/t $CO_2e$ and the 45Z duration is less than 2 years. The P15 pathway is the most competitive option when the LCFS credit is between $25/t $CO_2e$ and $75/t $CO_2e$ and the 45Z duration is less than 2 years. The P11 pathway outcompetes all others in the rest of the variable space. When the D5 RIN value is $0.75 and the D3 RIN is $2.25 (upper right panel in 3B), the P12 pathway is the most competitive one when the LCFS credit is below $25/t$CO_2e$ and the 45Z duration is less than 1 year. The P13 and P11 pathways become the dominant options for the rest of the parameter space (longer 45Z duration and higher LCFS favors P11). When the D5 RIN value is $1.5 and the D3 RIN is $2 (lower left panel in 3B), P15 pathway is the most competitive option when the LCFS credit is below $100/t$CO_2e$ and the 45Z duration is less than 3 years, and the rest of the parameter space is again dominated by P13 and P11. Finally, with D5 RIN at $1.5 and the D3 RIN at $3 (lower right panel in 3B), P12 is the most cost-effective option when subsidies from 45Z and LCFS are small, but P13 outperforms

all others across the rest of the parameter space. A wider range of D5 and D3 RIN values is explored in Figure S4.

**Measuring IRA incentives against the Social Cost of Carbon.** IRA incentives can be viewed as subsidies for achieving GHG emission reductions. In this context, it is of interest to express the incentives for each clean-energy pathway in terms of the Levelized Subsidy for Carbon Mitigation (LSCM, Eqn (4)) and compare these with recent U.S. federal government estimates of the Social Cost of Carbon (SCC) for the 2030 to 2040 timeframe [42]. The SCC is an estimate of the economic damages anticipated from climate change due to one additional tonne of $CO_2$ emissions. For this comparison, the amount of $CO_2$ mitigated by a pathway is taken to be the difference between the lifecycle emissions for the pathway and that for the pathway's fossil benchmark. Figure 4 shows LSCM values for all clean-energy pathways, along with the range in estimated SSC for 2030-2040. For the SLF pathways, we assume the 45Z credit is extended and available for the full assumed asset life (15 years), representing a perhaps optimistic case for liquid fuel pathways. Credits from RFS and LCFS have not been considered in determining these LSCM values.

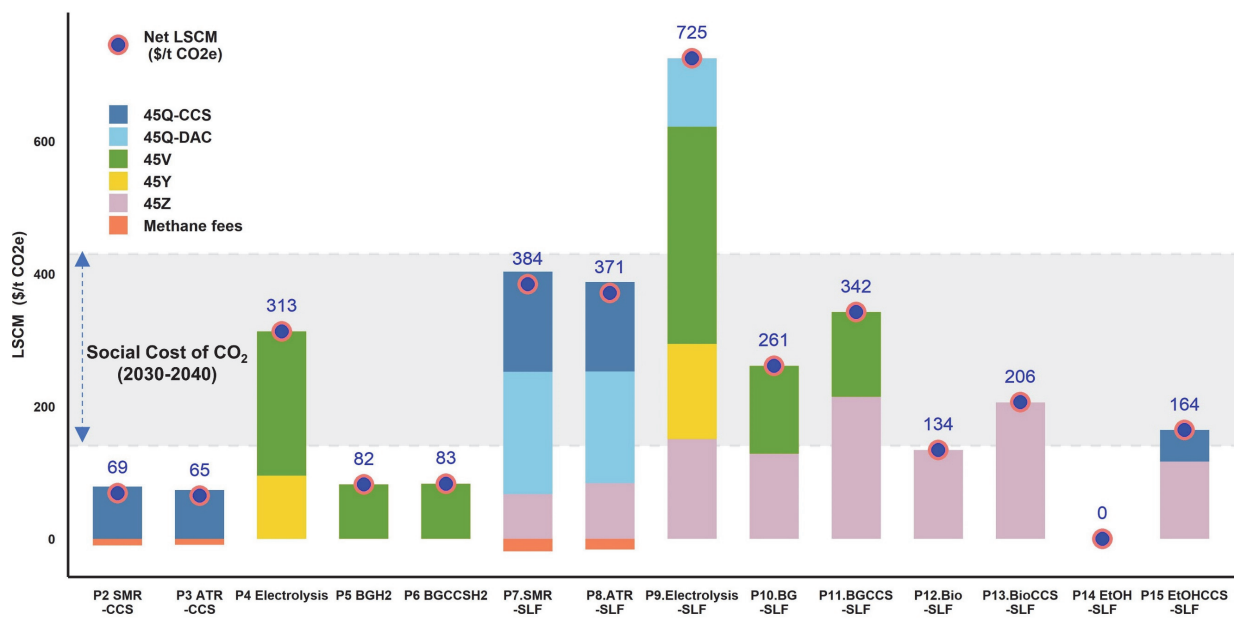

Figure 4. Levelized subsidy for carbon mitigation (LSCM) implied in IRA credits for all evaluated pathways. The assumed duration of 45Q, 45V/45Y, and 45Z/methane penalties are 12, 10, and 15 years, respectively. LSCM is the sum of IRA credits ($/kg $H_2$ or $/gal SLF) divided by the emissions avoided relative to fossil-derived hydrogen or liquid fuel. See Eqn (4). P14 does not receive any IRA credit. The indicated range in Social Cost of $CO_2$ emissions (SCC) is for discount rates from 1.5% to 2.5% [42]. For 2030 and 2040, the SCC ranges are $140 to 380/t $CO_2$ and $170 to 430/t $CO_2$, respectively.

LSCM values for four of the hydrogen pathways (P2, P3, P5, P6) are below the SCC range, and for one pathway (P4) it is in the mid-range of SCC values. Our earlier analysis (Figure 1A) showed that, with IRA credits, blue $H_2$ (P2 and P3) and green $H_2$ (P4) are cost-competitive with incumbent (gray) $H_2$. Figure 4 shows that their deployment, especially blue $H_2$, would be societally cost-effective as well, considering avoided emissions related damages. Our earlier analysis also showed that IRA credits for P5 and P6 pathways are insufficient to make these cost competitive with gray $H_2$. Figure 4 indicates that higher credits are justified from a societal perspective. For example, if 45Q credits were allowed for biogenic $CO_2$ sequestration with P6, as mentioned in our discussion of Figure 1B, the P6 pathway would be cost-competitive with gray hydrogen and the implied LSCM ($130/t $CO_2$e) would still be below the lower bound of SCC.

Figure 4 shows that LSCM values for SLF pathways fall within or below the SSC range, except for the electro-fuel pathway (P9). Our earlier analysis (Figure 3A) showed that, with IRA credits (assuming a 15-year 45Z duration), the P9, P11, P13 and P15 pathways would be cost-competitive with the fossil liquid-fuel benchmark. Figure 4 shows that deployment in the case of P11, P13 and P15 pathways also would be cost-competitive from a societal perspective, but not in the case of P9, for which the LSCM is far above the SCC range. Furthermore, we previously suggested that it might be justified to incentivize negative-emissions by allowing the pathway P6 to claim 45V credits for low-carbon hydrogen production and, additionally, 45Q credits for negative (biogenic) $CO_2$ emissions. It would not be appropriate for a similar negative-emissions incentive to be applied for liquid fuels production, however, because 45Z credits are already linked to a pathway's carbon intensity, such that the credit increases as the emissions grow more negative (Figure S2). Thus, allowing a SLF pathway like P11 or P13 to earn 45Q credits and 45Z credits simultaneously would, effectively, be providing two incentives for the same negative emissions.

**Discussion:**

Our results provide insights on each of the research questions posed in the introduction.

First, our analysis demonstrates that IRA subsidies will enable clean hydrogen to be cost-competitive with gray hydrogen. Blue hydrogen will have costs comparable to gray hydrogen thanks to the 45Q credit for CCS. Green hydrogen receives the highest subsidies among all evaluated hydrogen pathways due to the combination of a 10-year 45Y renewable electricity PTC and a 10-year 45V hydrogen PTC, which make electrolysis the lowest cost option for hydrogen production. Biomass-derived hydrogen, however, is not cost-competitive despite IRA subsidies. In our analysis, we used attributional LCA [43] and determined that green hydrogen had zero GHG emissions (Figure 1A) when carbon-free electricity sources are used. However, green hydrogen may result in consequential carbon emissions, as when procurement of carbon-free electricity for hydrogen production leads other electricity users to switch to fossil generated power to meet their demands [44]. This raises concerns regarding the environmental benefits of green hydrogen, at least until the grid is fully decarbonized. Further efforts on LCA methodology (e.g., attributional/consequential LCA selection, co-product accounting, and system boundary definition) are necessary to ensure electrolytic hydrogen has near-zero lifecycle emissions.

Second, our results indicate that IRA incentives favor the electrolytic route to hydrogen over others, because subsidies for different hydrogen production pathways are observed to be approximately proportional to their GHG emissions reductions ($76 /t $CO_2$e reduced), while the electrolysis route receives a disproportionately higher subsidy ($217/t $CO_2$e). This suggests the IRA is intended to incentivize technology innovation as well as emissions reductions in the case of green hydrogen. A similar technology policy rationale could also reasonably apply to biomass gasification routes to hydrogen, but this is not the case under current IRA provisions.

Third, for SLF pathways, we find that renewal and extension of the 45Z credit would be necessary for most low- and negative-carbon emissions SLF production pathways to be competitive with petroleum-derived jet fuels. The duration of a 45Z extension of 5 years would allow four clean SLF routes to achieve LCOFs that fall within the recent historical range in U.S. wholesale jet fuel prices, without considering the impact of any additional incentives (e.g., RFS or LCFS).

We also find that stacking 45V and 45Z credits by artificially separating production facilities, e.g., P10 and P11, provides for greater subsidies and results in a lower LCOF for SLF production from biomass than with standalone integrated facilities (P12 and P13), despite the higher capital costs and lower conversion efficiencies of biomass to fuel at separated production facilities. We further note that the IRA defines 45Z incentives to be a function of a pathway's life cycle GHG emissions. Accordingly, pathways with negative emissions receive the highest credits per gallon of fuel produced. This raises the possibility that 45Z credits by themselves may incentivize less energy-efficient (P11) over more energy-efficient (P13) bio-SLF production with CCS, because the lower efficiency enables greater $CO_2$ capture and storage to achieve stronger negative emissions. When federal RFS credits (RINs), which are available only to stand-alone biomass-SLF facilities (P12 and P13 pathways), are stacked with 45Z credits, the P13 pathway could be sufficiently incentivized that it becomes more competitive than the P11 pathway. In addition, LCFS credits can be stacked with IRA credits for any of the SLF pathways. LCFS and RFS credits have historically fluctuated over time (Figure S1), and there is uncertainty around what they will be in the 2030s. For a wide array of combinations of 45Z duration, LCFS credit level, and RIN value, our analysis finds that the P11, P12, P13 or P15 pathway is the most cost-competitive one against the median monthly-average wholesale price of jet fuel over the past decade. No SLF pathway is competitive with $2.2/gallon jet fuel for short 45Z durations combined with low LCFS and RIN values.

Finally, we find that IRA incentives for each pathway, when expressed in terms of LSCM, are below or within the range in the SCC estimated for 2030 to 2040, except for the electro-fuel pathway (P9), which receives a significantly higher total incentive due to stacking of multiple IRA credits. It is notable that, with IRA credits, BGCCSH2 (P6) is not cost-competitive with gray $H_2$ (P1), but has a LSCM well below the lower bound of the SCC range. If the IRA were to allow BGCCSH2 to simultaneously claim 45V credits for low-carbon hydrogen and 45Q credits for negative emissions, it would become cost-competitive with gray $H_2$, and the LSCM would still be below the lower bound of the SCC range.

In addition to the hydrogen and SLF pathways investigated in our analysis, one can envision other pathways to exploit stacking of credits to generate outsize financial benefits, but with marginal environmental gains. For example, 45Y can be applied to carbon-free electricity generators to power electrolysis, 45V then provides $3/kg $H_2$ to the electrolysis facility for its hydrogen production, and this hydrogen can end up in a hydrogen-fueled generator that could potentially claim a 45Y credit for its clean power generation. The system in this case intakes and produces carbon-free electricity, but the interior energy losses may be greater than another technology that performs the same service (e.g., battery electricity storage). For these reasons, policies that prioritize the use of fuels (hydrogen in particular) in sectors where there are no other alternatives or where decarbonization may be more challenging would be useful.

In general, we find that IRA achieves the objective of creating early commercial-scale deployment opportunities for emerging low-carbon hydrogen and liquid fuels technologies and pathways. Both green $H_2$ and blue $H_2$ pathways appear to be commercially viable under IRA incentives, while only modest extension of the 45Z clean fuels credit and/or incentives provided by the federal RFS or California LCFS make one or more SLF pathways cost competitive with historical fossil jet fuel prices. Accelerating market deployment of these clean fuels may not only contribute to near-term GHG emissions reductions, but also drive cost declines that enable deeper emissions reductions in the 2030s and 2040s both in the U.S. and around the world. However, we find the design of IRA tax incentives is likely to favor other pathways over those with the highest carbon mitigation potentials ($H_2$ from biomass gasification w/CCS and SLF from integrated biogasifier-FTS w/CCS), pathways that could prove to be critical options for achieving a net-zero emissions future.

**Limitations:**

Our work here is not without limitations, and we identify five key ones here that would benefit from additional examination in future work. (1) We have assumed that tax credits accrued throughout the supply chain will all be passed on to the final product (e.g., $H_2$ or SLF). For instance, in a project incorporating renewable generators for power generation and electrolyzers for $H_2$ manufacture, we assume that the full 45Y PTC value will go toward reducing the cost of electricity to the $H_2$ producer. In reality, the price of electricity to the $H_2$ producer will depend on competitive market dynamics or negotiated power purchase price. This could mean higher electricity costs to the $H_2$ producer. Liquid fuel pathways, especially eletrofuel, may be more complicated due to stacking of credits from multiple facilities. Although some entities, e.g., non-profit organizations, and projects claiming the 45Q credits (within the first 5 years of operation) are eligible to receive the IRA credits as "direct pay", i.e., they receive payments equivalent to the tax credits, a more common scenario may be to transfer to third-parties at some discount from full value. (2) We assume biogenic carbon to be carbon neutral. This generally holds for agricultural residues since they absorb $CO_2$ as they grow on an annual cycle, and they do not introduce new $CO_2$ into atmosphere when they are combusted. However, it may not apply for forest-derived biomass, which regrows on longer cycles. (3) For liquid fuel pathways we have evaluated Fischer-Tropsch and ethanol-to-SLF routes, but there are others that could monetize IRA credits and these would need to be evaluated to provide a more complete understanding of possibilities. (4) We did not differentiate our analysis of $H_2$ pathways by intended use of the $H_2$. In particular, we did not explicitly assess transportation use of $H_2$, which would likely require substantial new delivery and dispensing infrastructure to be established, but which might also qualify for LCFS and/or RFS credits. (5) IRA subsidies may very well alter the international competitiveness of clean $H_2$ and SLF produced in the U.S. We have evaluated chosen to examine only domestic competitiveness implications, but consideration of potential international implications would also be of interest.

**Supporting Information:** Detailed summary and interpretation of Inflation Reduction Act (IRA), Renewable Fuel Standard (RFS) and California's Low Carbon Fuel Standard (LCFS) incentives; detailed summary of evaluated technologies; historical fuels and incentives prices; additional discussion of extending 45Z and complete design space of RIN credits; and comparisons of IRA credits and Social Cost of Carbon (SCC).

**Acknowledgments:**

Funding for this work was provided by Princeton University's Carbon Mitigation Initiative (BP funded), Andlinger Center for Energy and the Environment, and Zero-Carbon Technology Consortium (funded by gifts from GE, Google, ClearPath Foundation, and Breakthrough Energy).

**Declaration of Interests:**

J.D.J. is part owner of DeSolve, LLC, which provides techno-economic analysis and decision support for clean energy technology ventures and investors. A list of clients can be found at https://www.linkedin.com/in/jessedjenkins. He serves on the advisory boards of Eavor Technologies Inc., a closed-loop geothermal technology company, and Rondo Energy, a provider of high-temperature thermal energy storage and industrial decarbonization solutions, and has an equity interest in both companies. He also provides policy advisory services to Clean Air Task Force, a non-profit environmental advocacy group, and serves as a technical advisor to MUUS Climate Partners and Energy Impact Partners, both investors in early-stage climate technology companies. E.D.L. is on the advisory board and has an equity interest in DG Fuels, an emerging renewable hydrogen and biogenic synfuels company.

**Author contributions:**

**Fangwei Cheng:** Conceptualization, Investigation, Methodology, Formal analysis, Visualization, Writing – original draft; **Hongxi Luo:** Investigation, Formal analysis, Visualization, Writing – review & editing; **Jesse Jenkins**: Conceptualization, Writing – review & editing, Supervision, Funding acquisition; **Eric Larson**: Conceptualization, Investigation, Writing – review& editing, Supervision, Funding acquisition, Project administration

**References:**

1. White House, FACT SHEET: President Biden Sets 2030 Greenhouse Gas Pollution Reduction Target Aimed at Creating Good-Paying Union Jobs and Securing U.S. Leadership on Clean Energy Technologies. 2021. https://www.whitehouse.gov/briefing-room/statements-releases/2021/04/22/fact-sheet-president-biden-sets-2030-greenhouse-gas-pollution-reduction-target-aimed-at-creating-good-

paying-union-jobs-and-securing-u-s-leadership-on-clean-energy-technologies/ (accessed Jan 16th, 2023).
2.	White House, Executive Order on Tackling the Climate Crisis at Home and Abroad. 2021. https://www.whitehouse.gov/briefing-room/presidential-actions/2021/01/27/executive-order-on-tackling-the-climate-crisis-at-home-and-abroad/ (accessed Jan 16th, 2023).
3.	White House, The long-term strategy of the United States: pathways to net-zero greenhouse gas emissions by 2050. 2021. https://www.whitehouse.gov/wp-content/uploads/2021/10/US-Long-Term-Strategy.pdf (accessed March 17th, 2023).
4.	Senate Democrats, Summary: The inflation Reduction Act of 2022. 2022. https://www.democrats.senate.gov/imo/media/doc/inflation_reduction_act_one_page_summary.pdf (accessed Dec 14th, 2022).
5.	Jenkins, J. D., Farbes, Jamil, Jones, Ryan, & Mayfield, Erin N., REPEAT Project Section-by-Section Summary of Energy and Climate Policies in the 117th Congress [Data set] 2022. https://doi.org/10.5281/zenodo.6993118.
6.	Ramseur, J. L., Inflation Reduction Act Methane Emissions Charge: In Brief. Congressional Research Service. https://crsreports.congress.gov/product/pdf/R/R47206 (accessed June 24th, 2023).
7.	Jenkins, J. D.; Mayfield, E. N.; Farbes, J.; Jones, R.; Patankar, N.; Xu, Q. *Preliminary Report: The Climate and Energy Impacts of the Inflation Reduction Act of 2022*; 2022.
8.	Larsen, J.; King, B.; Kolus, H.; Dasari, N.; Hiltbrand, G.; Herndon, W. *A Turning Point for US Climate Progress: Assessing the Climate and Clean Energy Provisions in the Inflation Reduction Act*; Rhodium Group: 2022.
9.	Mahajan, M.; Ashmoore, O.; Rissman, J.; Orvis, R.; Gopal, A. Modeling the Inflation Reduction Act Using the Energy Policy Simulator. *Energy Innovation: Policy Technology.* **2022**.
10.	Bistline, J.; Blanford, G.; Brown, M.; Burtraw, D.; Domeshek, M.; Farbes, J.; Fawcett, A.; Hamilton, A.; Jenkins, J.; Jones, R.; King, B.; Kolus, H.; Larsen, J.; Levin, A.; Mahajan, M.; Marcy, C.; Mayfield, E.; McFarland, J.; McJeon, H.; Orvis, R.; Patankar, N.; Rennert, K.; Roney, C.; Roy, N.; Schivley, G.; Steinberg, D.; Victor, N.; Wenzel, S.; Weyant, J.; Wiser, R.; Yuan, M.; Zhao, A. Emissions and energy impacts of the Inflation Reduction Act. *Science.* **2023**, *380* (6652), 1324-1327.
11.	Williams, J. H.; Jones, R. A.; Haley, B.; Kwok, G.; Hargreaves, J.; Farbes, J.; Torn, M. S. Carbon‐neutral pathways for the United States. *AGU Advances.* **2021**, *2* (1), e2020AV000284.
12.	Bouckaert, S.; Pales, A. F.; McGlade, C.; Remme, U.; Wanner, B.; Varro, L.; D'Ambrosio, D.; Spencer, T. Net Zero by 2050: A Roadmap for the Global Energy Sector. **2021**.
13.	Larson, E.; C. Greig; J. Jenkins; E. Mayfield; A. Pascale; C. Zhang; J. Drossman; R. Williams; S. Pacala; R. Socolow; EJ Baik; R. Birdsey; R. Duke; R. Jones; B. Haley; E. Leslie; Paustian, K.; Swan, A. *Net-Zero America: Potential Pathways, Infrastructure, and Impacts, Final report*; Princeton University, Princeton, NJ, 29 October 2021.
14.	US Department of Energy, US National Clean Hydrogen Strategy and Roadmap. 2022. https://www.hydrogen.energy.gov/pdfs/us-national-clean-hydrogen-strategy-roadmap.pdf (accessed June 28th, 2023).
15.	Fan, Z.; Friedmann, S. J. Low-carbon production of iron and steel: Technology options, economic assessment, and policy. *Joule.* **2021**, *5* (4), 829-862.
16.	He, G.; Mallapragada, D. S.; Bose, A.; Heuberger-Austin, C. F.; Gençer, E. Sector coupling via hydrogen to lower the cost of energy system decarbonization. *Energy & Environmental Science.* **2021**, *14* (9), 4635-4646.
17.	Muratori, M.; Kunz, T.; Hula, A.; Freedberg, M. *US National Blueprint for Transportation Decarbonization: A Joint Strategy to Transform Transportation*; United States. Department of Energy. Office of Energy Efficiency and …: 2023.

18. Griffiths, S.; Sovacool, B. K.; Kim, J.; Bazilian, M.; Uratani, J. M. Industrial decarbonization via hydrogen: A critical and systematic review of developments, socio-technical systems and policy options. *Energy Research & Social Science.* **2021,** *80*, 102208.
19. Spath, P. L.; Mann, M. K. *Life cycle assessment of hydrogen production via natural gas steam reforming*; National Renewable Energy Lab.(NREL), Golden, CO (United States): 2000.
20. U.S. Energy Information Adminstration, Energy use for transportation. https://www.eia.gov/energyexplained/use-of-energy/transportation.php (accessed Jan, 10th, 2022).
21. Ibenholt, K. Explaining learning curves for wind power. *Energy Policy.* **2002,** *30* (13), 1181-1189.
22. Cheng, F.; Luo, H.; Jenkins, J. D.; Larson, E. D. The value of low-and negative-carbon fuels in the transition to net-zero emission economies: Lifecycle greenhouse gas emissions and cost assessments across multiple fuel types. *Applied Energy.* **2023,** *331*, 120388.
23. U.S. Environmental Protection Agency, Renewable Fuel Standard Program. https://www.epa.gov/renewable-fuel-standard-program (accessed Dec 14th, 2022).
24. California Air Resource Board, Low Carbon Fuel Standard. https://ww2.arb.ca.gov/our-work/programs/low-carbon-fuel-standard (accessed Dec 14th, 2022).
25. Environmental Protection Agency, RIN Trades and Price Information. https://www.epa.gov/fuels-registration-reporting-and-compliance-help/rin-trades-and-price-information (accessed Dec 15th, 2022).
26. California Air Resource Board, Monthly LCFS Credit Transfer Activity Reports. https://ww2.arb.ca.gov/resources/documents/monthly-lcfs-credit-transfer-activity-reports (accessed Dec 15th, 2022).
27. California Air Resource Board, LCFS Electricity and Hydrogen Provisions. https://ww2.arb.ca.gov/resources/documents/lcfs-electricity-and-hydrogen-provisions (accessed Dec 14th, 2022).
28. Littlefield, J. A.; Marriott, J.; Schivley, G. A.; Skone, T. J. Synthesis of recent ground-level methane emission measurements from the U.S. natural gas supply chain. *Journal of Cleaner Production.* **2017,** *148*, 118-126.
29. Canter, C. E.; Qin, Z.; Cai, H.; Dunn, J. B.; Wang, M.; Scott, D. A. Fossil energy consumption and greenhouse gas emissions, including soil carbon effects, of producing agriculture and forestry feedstocks. *2016 Billion‐Ton Report: Advancing Domestic Resources for a Thriving Bioeconomy Volume 2: Environmental Sustainability Effects of Select Scenarios from Volume 1 ORNL/TM‐2016/727.* **2017**, 85-137.
30. Kreutz, T. G.; Larson, E. D.; Elsido, C.; Martelli, E.; Greig, C.; Williams, R. H. Techno-economic prospects for producing Fischer-Tropsch jet fuel and electricity from lignite and woody biomass with CO2 capture for EOR. *Applied Energy.* **2020,** *279*, 115841.
31. Wang, M.; Elgowainy, A.; Lu, Z.; Baek, K.; Bafana, A.; Benavides, P.; Burnham, A.; Cai, H.; Cappello, V.; Chen, P.; Gan, Y.; Gracida-Alvarez, U.; Hawkins, T.; Iyer, R.; Kelly, J.; Kim, T.; Kumar, S.; Kwon, H.; Lee, K.; Lee, U.; Liu, X.; Masum, F.; Ng, C.; Ou, L.; Reddi, K.; Siddique, N.; Sun, P.; Vyawahare, P.; Xu, H.; Zaimes, G. *Greenhouse gases, Regulated Emissions, and Energy use in Technologies Model ® (2022 .Net)*, United States, 2022.
32. Greig, C.; Larson, E.; Kreutz, T.; Meerman, J.; Williams, R. *Lignite-plus-Biomass to Synthetic Jet Fuel with CO2 Capture and Storage: Design, Cost, and Greenhouse Gas Emissions Analysis for a Near-Term First-of-a-Kind Demonstration Project and Prospective Future Commercial Plants*; United States, 2017, available from https://doi.org/10.2172/1438250, 2017-09-01.
33. Cummins Inc, HyLYZER®WATER ELECTROLYZERS. https://www.cummins.com/sites/default/files/2021-08/cummins-hylyzer-1000-specsheet.pdf (accessed July 10th, 2023).

34. Duong, C.;  Bower, C.;  Hume, K.;  Rock, L.; Tessarolo, S. Quest carbon capture and storage offset project: Findings and learnings from 1st reporting period. *International Journal of Greenhouse Gas Control.* **2019,** *89*, 65-75.
35. Air Products, Louisiana Clean Energy Complex. https://www.airproducts.com/campaigns/la-blue-hydrogen-project (accessed July 9th,  2023).
36. Malhotra, A.; Schmidt, T. S. Accelerating Low-Carbon Innovation. *Joule.* **2020,** *4* (11), 2259-2267.
37. Mesfun, S. A., Biomass to liquids (BtL) via Fischer-Tropsch – a brief review. European Technology and Innovation Platform – Bioenergy, January 2021. https://www.etipbioenergy.eu/images/ETIP_B_Factsheet_BtL_2021.pdf.
38. Geleynse, S.;  Brandt, K.;  Garcia‐Perez, M.;  Wolcott, M.; Zhang, X. The alcohol‐to‐jet conversion pathway for drop‐in biofuels: techno‐economic evaluation. *ChemSusChem.* **2018,** *11* (21), 3728-3741.
39. World Economic Forum, Aviation commitment. 2022. https://www3.weforum.org/docs/WEF_FMC_Aviation_2022.pdf (accessed March 17th,  2023).
40. Xu, H.;  Lee, U.; Wang, M. Life‐cycle greenhouse gas emissions reduction potential for corn ethanol refining in the USA. *Biofuels, Bioproducts and Biorefining.* **2022,** *16* (3), 671-681.
41. U.S. Energy Information Adminstration, U.S. Kerosene-Type Jet Fuel Wholesale/Resale Price by Refiners. https://www.eia.gov/dnav/pet/hist/LeafHandler.ashx?n=pet&s=ema_epjk_pwg_nus_dpg&f=m (accessed Dec 15th,  2022).
42. U.S. Environmental Protection Agency, Supplementary Material for the Regulatory Impact Analysis for the Supplemental Proposed Rulemaking, "Standards of Performance for New, Reconstructed, and Modified Sources and Emissions Guidelines for Existing Sources: Oil and Natural Gas Sector Climate Review". 2022. https://www.epa.gov/system/files/documents/2022-11/epa_scghg_report_draft_0.pdf.
43. Ekvall, T., Attributional and consequential life cycle assessment. In *Sustainability Assessment at the 21st century*, IntechOpen: 2019.
44. Ricks, W.;  Xu, Q.; Jenkins, J. D. Minimizing emissions from grid-based hydrogen production in the United States. *Environmental Research Letters.* **2023,** *18* (1), 014025.

# Supplementary Information for

# Inflation Reduction Act impacts on the economics of clean hydrogen and synthetic liquid fuels

Fangwei Cheng[1*], Hongxi Luo[1], Jesse D. Jenkins[1,2], and Eric D. Larson[1]

1. Andlinger Center for Energy and the Environment, Princeton University, Princeton, NJ, USA

2. Department of Mechanical and Aerospace Engineering, Princeton University, Princeton, NJ, USA

* Corresponding author: fangweic@princeton.edu

# Contents

## 1. Summary of applicable incentives provide by the IRA

Table S1 summarizes policies from the IRA that are applicable for $H_2$ and SLF production, including carbon sequestration or utilization credits (45Q), clean hydrogen credits (45V), sustainable aviation fuel and clean fuel credits (40B/45Z), clean electricity credits (45Y), and methane emission penalties. Facilities in our analysis are assumed to enter service by 2032 and operate for 15 years. We consider credit durations of 12 years for the 45Q PTC, 10 years for the 45V and 45Y PTC, and 15 years for the methane emission fee, as stipulated in the IRA. The 45Z credit is stipulated in the IRA to expire in 2027. We explore, through sensitivity analysis in our paper, possible impacts of this credit being extended beyond 2030.

Table S1. Summary of applicable policies in the Inflation Reduction Act (2022 USD) [1]

| Policies | Summary | Term |
|---|---|---|
| **Carbon oxide sequestration credit (45Q)** | $85/t $CO_2$ tax credit for $CO_2$ sequestration<br><br>$60/t $CO_2$ tax credit for $CO_2$ utilization<br><br>$180/t $CO_2$ tax credit for DAC $CO_2$ sequestration<br><br>$130/t $CO_2$ tax credit for DAC $CO_2$ utilization | Commence construction: 2023 – 2032<br><br>Subsidy term: first 12 years of facility operations.<br><br>Inflation adjustment: Credit will be inflation-adjusted beginning in 2027 and indexed to base year 2025 |
| **Clean hydrogen production credit (45V)** | $0.6/kg $H_2$ if life cycle GHG emissions >= 2.5 kg $CO_2e/H_2$ & < 4 kg $CO_2e/H_2$<br><br>$0.75/kg $H_2$ if life cycle GHG emissions >= 1.5 kg $CO_2e/H_2$ & < 2.5 kg $CO_2e/H_2$<br><br>$1.002/kg $H_2$ if life cycle GHG emissions >= 0.45 kg $CO_2e/H_2$ & < 1.5 kg $CO_2e/H_2$<br><br>$3/kg $H_2$ if life cycle GHG emissions < 0.45 kg $CO_2e/H_2$ | Commence construction: 2023 - 2032<br><br>Subsidy term: first 10 years of facility operations.<br><br>Inflation adjustment: The credit will be adjusted annually for inflation and indexed to base year 2022 |
| **Sustainable aviation fuel credit (40B)** | $1.25/gal for sustainable aviation fuel (SAF) if 50% emission reduction is achieved as compared to petroleum-based jet fuel + 1 cent/percent after 50% up to 50 cents total supplement credit, i.e., maximum $1.75/gal if zero/negative emissions are achieved. | Commence construction: n/a<br><br>Subsidy term: available for operations during 2023-2024.<br><br>Inflation adjustment: The credit will be adjusted annually for inflation and indexed to base year 2022 |
| **Clean fuel production credit (45Z)** | Base credit is $0.2/gal ($0.35/gal for SAF) and full value bonus credit is $1/gal ($1.75/gal for SAF), actual credit value is based on "emission factor" times qualifying base or bonus rate. Emissions factor = (50 – carbon intensity of fuel)/50. The carbon intensity of fuel is represented as $kgCO_2e/MMBTU_{LHV}$ | Commence construction: n/a<br><br>Subsidy term: available for operations during 2025-2027.<br><br>Inflation adjustment: The credit will be adjusted annually for inflation and indexed to base year 2022 |
| **Clean Electricity Production Credit (45Y)** | 10-year PTC worth $26/MWh (in 2022 USD, inflation-adjusted thereafter) | Commence construction: 100% value for projects that commence construction prior to 2033 or the year after annual greenhouse gas emissions from electricity production in the U.S. is <= 25% of 2022 emissions levels, whichever comes later; 75% value for projects that commence construction in a subsequent year; 50% value for projects that commence construction in year after that.<br><br>Subsidy term: first 10 years of facility operations<br><br>Inflation adjustment: The credit will be adjusted annually for inflation and indexed to base year 2022 |
| **Methane emissions reduction program** | $900/tonne methane fee in 2023, ramping to $1200/tonne methane in 2024 and $1500/tonne methane in 2025 and thereafter. | Starts in 2023. |

## 2. Renewable Fuel Standard and California Low-Carbon-Fuel Standard incentives

In addition to provisions in the IRA, low-carbon and negative-carbon emission fuels used in the transportation sector may be eligible to receive incentives under the federal Renewable Fuels Standard (RFS) and California's Low Carbon Fuel Standard (LCFS).

RFS is a national policy that mandates the blending of renewable fuel to reduce or replace fossil-derived transportation fuel, heating oil, or jet fuel [2]. Under RFS, approved pathways are assigned a Renewable Fuel Category ("D" code) based on the type of fuel, feedstock, and production process. Currently, four categories of renewable fuels are certified under the RFS: cellulosic biofuel (D3 or D7), biomass-based diesel (D4), advanced biofuel (D5), and renewable fuel (D6). The D-code represents the type of Renewable Identification Number (RIN) or the numbered credit that the fuel is eligible for. A RIN credit is assigned to an eligible fuel for each volume of that fuel having energy content equal to that of one gallon of ethanol. Qualifying fuels with higher volumetric energy contents than ethanol will generate more than 1 RIN per gallon. For example, the LHV energy content of ethanol and the synthetic liquid fuel (SLF) considered in the analysis in our paper are 0.077 million BTU/gallon and 0.126 million BTU/gallon, respectively. This results in a 1.64 equivalence ratio, meaning that one gallon of synthetic fuel is eligible to receive 1.64 RIN credits (Eqn S1).

$$RFS_p (\$/gal) = 1.64 \times \text{RIN credit} (\$)/1 \text{ gallon SLF} \tag{S1}$$

RINs are generated by fuel producers who use biomass as feedstock. RINs can be traded freely in the market, where obligated parties, such as non-renewable fuel refiners or importers, will purchase and retire RINs to fulfill their Renewable Volume Obligation (RVO). The RVO is the federally mandated percentage renewable content of a fuel supplier's fuel. Historical RIN prices for D3 – D6 are given in Figure S1(C). SLF produced via pathways P12 and P13 may claim either cellulosic biofuel (D3) RINs or advanced biofuel (D5) RINs. As for P14 and P15, we assume P14 and P15 can claim D6 and D5 RINs, respectively. Due to ongoing insufficiency in the capacity for producing cellulosic biofuels in the U.S., the EPA has been issuing waiver credits such that obligated parties can fulfill their cellulosic biofuel blending mandate by purchasing cellulosic waiver credits (CWC) from EPA together with advanced biofuel (D5) RINs. Hence, the price difference between the D5 RINs and D3 RINs is affected by both the CWC price and the market dynamics. During 2012 to 2022, this price difference (monthly average) varies between $0 and $1.98, with 25[th] percentile at $0.49 and 75[th] percentile at $1.46. In the analysis of this work, we assume that the price of D3 RINs is $0.5 to $1.5 plus the price of D5 RINs.

The LCFS in California (and other states like Washington and Oregon) aims to reduce GHG emissions in the transportation sector by providing credits for low-carbon emission fuels, such as biomass-derived liquid fuels, hydrogen, electricity, and renewable natural gas [3]. A fuel supplier earns LCFS credits if the carbon intensity (CI) of its fuels is less than the benchmark mandated by the state. The SLF analyzed in our studies contains 82% SPK, thus we approximate the SLF as a substitute for jet fuel when determining LCFS credits. California's benchmark CI for jet fuel was 89.37 kg $CO_2$e/$GJ_{LHV}$ in 2022 and will be lowered to 80.36 kg $CO_2$e/$GJ_{LHV}$ by 2030 [4]. A post-2030 schedule of further reductions in the benchmark has not been published, so we assume the benchmark will remain at the 2030 level for pathways analyzed in our paper. The LCFS credits for SLF pathways are determined by the following equation:

$$LCFS_p (\$/gal) = (CI_{standard}^{jet\ fuel} - CI_p^{SLF}) \times \frac{0.126\ GJ_{LHV}}{gal} \times \frac{1\ MT}{10^3 (kg\ CO2e)} \times \text{Credit Price} (\$/t\ CO2) \tag{S2}$$

where $LCFS_p$ ($/gal) is the carbon credit received by a SLF pathway (p); $CI_{standard}^{jet\ fuel}$ (kg/$GJ_{LHV}$) is the state CI benchmark for jet fuel in a given year; $CI_p^{SLF}$ (kg/$GJ_{LHV}$) is the CI of a SLF pathway (p) in our

analysis; $\frac{0.126\ GJ_{LHV}}{gal}$ converts $GJ_{LHV}$ into gallons, and *Credit Price* is the LCFS credit price ($/t $CO_2$e). Historical LCFS credit prices are given in Figure S1(D).

## 3. Summary of evaluated technologies for $H_2$ and synthetic liquid fuel production

Assumed performance and cost characteristics for evaluated $H_2$ and SLF production technologies are shown in Table 2 in the main paper. Table S2 here provides more supporting details. Table S3 shows the assumed input energy and materials prices. These are based on Cheng et al [5], who evaluated six hydrogen production technologies, including steam methane reforming (SMR, the current benchmark), steam methane reforming with CCS (SMR-CCS), auto thermal reforming with CCS (ATR-CCS), biomass gasification (BG-$H_2$), and biomass gasification with CCS (BGCCS-$H_2$). The natural gas pathways incur the IRA's methane emissions penalties (Table S1) for upstream fugitive methane emissions. (The penalty is incorporated into the assumed natural gas input price, Table S3.) The SMR-CCS, ATR-CCS and BGCCS-$H_2$ pathways earn either the IRA's 45Q or 45V credit, whichever provides greater remuneration. BG-$H_2$ and electrolysis pathways get 45V credits. Additionally, a 45Y clean electricity credit is incorporated into the assumed price of electricity generated by solar or wind resources (Table S3). Clean hydrogen used in transport or for petroleum-fuel upgrading can be eligible for LCFS credits [6], but we have not considered this possibility here because the former would require taking into account expansion of infrastructure for $H_2$ delivery and dispensing [7], and the latter would lead to credits for refineries, rather than $H_2$ producers [8].

For SLF production pathways, Cheng et al [5] evaluated five pathways that include stand-alone facilities making $H_2$ for delivery to separate SLF production facilities. The SLF facilities use the $H_2$, together with carbon-neutral $CO_2$ inputs, in reverse-water-gas-shift (RWGS) and Fischer-Tropsch synthesis (FTS) processes. The RWGS-FTS pathways (P7 – P11) are assumed to use low-carbon or negative-carbon emissions hydrogen from each of the five aforementioned $H_2$ production pathways (excluding SMR) and carbon-neutral $CO_2$ of biogenic or captured directly from the air (DAC). Consistent with Cheng et al. [5], when carbon-neutral $CO_2$ is purchased from a DAC facility powered by renewable energy (pathways P7 – P9), the cost of $CO_2$ is 280 $/t $CO_2$ (2022 $). $CO_2$ utilization from DAC is credited with 130 $/t $CO_2$ under 45Q for the first 12 years of operation. For pathways P10 and P11, biogenic $CO_2$ is used as the carbon source for the RWGS-FTS at no cost beyond the cost for separating it from hydrogen that is intrinsic to the cost for bio-$H_2$ from pathways P5 and P6.

Here, we additionally evaluate four stand-alone SLF production pathways: one integrated biomass gasifier-FTS facility without CCS (pathway P12) and one with CCS (P13) and two ethanol-to-SNL pathways, one with venting of fermentation $CO_2$ from ethanol production and one with capture and storage of that $CO_2$. Performance and cost estimates consistent with other pathways evaluated here are developed for these four additional pathways (see Table S2, notes *g* and *h*).

All SLF pathways are assumed to produce sustainable aviation fuel (SAF) as the primary output, with quality that meets Annex A1 provisions of ASTM International Standard D165. The produced synthetic fuel is assumed to consist of 82% synthetic paraffinic kerosene (SPK) and 18% naphtha by energy content, based on detailed process designs of Kreutz et al. [9].

In addition to exploring the impact of 45Z credits extended into the 2030s, we also assume all SLF pathways could qualify for LCFS credits. (SAF is eligible under the LCFS program [10].) To date, only integrated BioSLF (P12), integrated BioCCS-SLF (P13), EtOH-SLF (P14), and EtOHCCS-SLF (P15) have been certified as eligible to receive RINs under the RFS, as cellulosic biofuel (D3), advanced biofuel (D5) or renewable biofuel (D6), and we explore the impact of RIN prices on the costs for SLF from these two pathways. Pathways P7 – P11, which produce SLF at stand-alone RWGS-FTS facilities, do not currently qualifying for credits, because they do not use biomass directly as feedstocks [11].

Table S2. Summary of evaluated hydrogen and SLF technologies.[a] Cost parameters are adjusted from the 2016$ values given in Cheng, *et al* [5] to 2022$, but not including the anomalous cost spikes seen in 2021 and 2022

as a result of Covid-related supply-chain bottlenecks.[b] Parameter values here are given per unit of output capacity or output energy. All fuel values are on a higher heating value basis.

| | Technology | Input Feedstock Intensity (IFI) ($GJ_{in}/GJ_{out}$) | Co-product[c] ($GJ/GJ_{H2}$) | $CO_2$ capture rate (%)[d] | CAPEX ($/kW capacity) | Fixed O&M ($/kW-year) | Variable O&M ($/GJ) |
|---|---|---|---|---|---|---|---|
| $H_2$ | SMR | 1.23 (NG) | 0 | 0 | 543 | 19 | 0.36 |
| | SMR-CCS | 1.26 (NG) | 0.024 | 91.2 | 943 | 28 | 0.00 |
| | ATR-CCS | 1.2 (NG) | 0.048 | 94.1 | 825 | 25 | 0.00 |
| | Electrolysis | 1.28 (electricity) | 0 | 0 | 1407 | 16 | 0.00 |
| | BG-H2 | 1.78 (biomass) | - 0.079 | 0 | 2482 | 44 | 3.89 |
| | BGCCS-H2 | 1.78 (biomass) | - 0.023 | 87 | 2587 | 46 | 3.89 |
| SLF | RWGS+FTS[e,f] | 1.47 ($H_2$) | 0 | 0 | 1004 | 27 | 0.73 |
| | Integrated BioFTS[f,g] | 2.02 (biomass) | 0 | 0 | 3826 | 77 | 5.15 |
| | Integrated BioFTS-CCS[f,g] | 2.02 (biomass) | 0 | 87 | 3944 | 79 | 5.15 |
| | Ethanol-to-Jet[f,h] | 1.09 (ethanol) | 0 | 0 | 258 | 26 | 2.15 |
| | Ethanol with CCS-to-Jet[f,h,i] | 1.09 (ethanol) | 0 | 0 | 258 | 26 | 2.15 |

(a) See Cheng et al [5] for detailed description of the input data used to produce values in this table, except for the Integrated BioFTS, Integrated BioFTS-CCS, Ethanol-to-Jet, and Ethanol with CCS-to-Jet values. See note (g) and (h) below for discussion of these four technologies.

(b) To express costs to 2022 $, the original 2016 $ costs of Cheng et al. [5] were multiplied by a ratio between the average value of annual Chemical Engineering Plant Cost Indexes (CEPCI) during 2010-2020 (576) and the annual CEPCI for 2016 (542) [12]. During 2010 - 2020, the CEPCI was relatively stable, fluctuating only $\pm 6\%$ from that decade's average. From 2020 to 2021, however, the increase in CEPCI averaged an unprecedently high 17% per year, reaching 817 for 2022. Our analysis assumes that as supply chains return to pre-Covid normalcy the anomalous cost increases seen in 2021 and 2022 will subside again to the more stable level observed in the 2010s decade.

(c) Co-product is electricity in all cases. Positive values are inputs and negative values are outputs.

(d) $CO_2$ capture rate is the amount of carbon captured divided by the amount of carbon available for capture. The latter is the carbon in the input feedstock less the carbon in the produced fuel. For the Ethanol with CCS-to-Jet pathway, the $CO_2$ capture is conducted at the ethanol production plant, not at the ethanol upgrading plant. Thus, the $CO_2$ capture rate is 0%.

(e) For the RWGS-FTS (Reverse-Water-Gas-Shift – Fischer-Tropsch-Synthesis) pathways, input of 67.7 kg of carbon-neutral $CO_2$ (from direct air capture or bio-derived) is required [5], in addition to 1.47 GJ of hydrogen, to produce 1GJ SLF. SLF production via RWGS+FTS is assessed for $H_2$ from each of the $H_2$ production routes in this table, except SMR.

(f) Produced SLF is assumed to contain 82% synthetic paraffinic kerosene (SPK) and 18% naphtha on an LHV energy basis. This is the composition by Kreutz et al. [9] based on detailed process simulations of biomass gasifier-Fischer-Tropsch synthesis. We assume an equivalent fuel composition can be achieved in ethanol conversion to SLF.

(g) These pathways involve stand-alone facilities that gasify biomass feedstocks to produce synthesis gas converted into SLF at the same facility. One pathway excludes CCS (integrated BioFTS) and the other includes CCS (integrated BioFTS-CCS). Neither of these pathways were included in Cheng et al. [5]. We developed performance and cost metrics for these that are consistent with those in Cheng et al. [5].

We estimate the IFI for integrated facilities as follows. First, note that in an integrated facility the $H_2$:CO molar ratio in the synthesis gas from the gasifier is adjusted via water-gas-shift (WGS) to the optimal ratio of 2:1 for FTS, after which the mixture passes to a Rectisol unit for acid gas removal, before it then passes to the FTS unit. For quantitative illustration purposes, consider an amount of biomass input to the facility that enables production of 300 total moles of CO + $H_2$ syngas for the FTS feed. 100 of these will be CO and 200 will be $H_2$.

Second, consider what happens when hydrogen production and FTS occur in separate facilities starting with the same initial amount of biomass, e.g. BGH2 + RWGS-FTS (pathway P10) or BGCCSH2 + RWGS-FTS (P11). In this case, the hydrogen production facility subjects the synthesis gas from the gasifier to a deep WGS step to maximize hydrogen content in the gas, which then undergoes pressure swing adsorption (PSA), where typically 80% of the feed $H_2$ is recovered at 99.99% purity to be sent to the RWGS-FTS facility. A molar $H_2$:$CO_2$ input ratio to the RWGS-FTS facility of 3:1 is required, since the production of each mole of CO from $CO_2$ via RWGS consumes one mole of $H_2$, leaving a $H_2$:CO ratio of 2:1 in the feed gas to the FTS unit. Again, for quantitative illustration, if we have the same biomass input rate as for the integrated facility, there will be 300 total moles of CO + $H_2$ available for deep WGS, after which there will be perhaps 10 moles of CO and 290 moles of $H_2$. Then, following PSA hydrogen separation, there would be 0.8*290 = 232 moles of $H_2$ available to send to the RWGS-FTS facility. One-third of this $H_2$ will be used to convert $CO_2$ to CO in the RWGS, resulting in 154.7 moles of $H_2$ and 77.3 moles of CO in the syngas feed to the FTS facility. These values are 77.3% of the $H_2$ and CO gas flows to the FTS unit in the integrated facility, so the IFI of the integrated facility, correspondingly, will be 77.3% of the IFI of the pathway with separate facilities:

$$IFI_{sep} = 1.78 \frac{GJ_{bio-in}}{GJ_{H2-out}} \cdot 1.47 \frac{GJ_{H2-in}}{GJ_{FTL-out}} = 2.62 \frac{GJ_{bio-in}}{GJ_{FTL-out}}, \text{ and } IFI_{int} = 2.62 * 0.773 = 2.02 \frac{GJ_{bio-in}}{GJ_{FTL-out}}.$$

We estimate unit capital cost (CAPEX) for the integrated facilities following the same logic as above. The CAPEX is the sum of two quantities: (1) the product of the CAPEX of the biomass-$H_2$ production facility (as in this table), the IFI of the RWGS-FTS facility

S-5

(1.47 $GJ_{H2-in}/GJ_{SLF-out}$) and the $IFI_{int}/IFI_{sep}$ ratio (0.773), and (2) the CAPEX of the RWGS-FTS facility (in this table). The FOM and VOM costs are estimated analogously. We neglect any co-products.

(h) To determine IFI for ethanol pathways, we adopted the mass-based input feedstock:product output ratio of 0.6 based on detailed process designs of Geleynse et al [13]:

$$IFI = \frac{HHV_{Ethanol} \times m_{Ethanol}}{HHV_{fuel} \times m_{fuel}} = \frac{29.7 \frac{MJ}{kg} \times 1\ kg}{45.5 \frac{MJ}{kg} \times 1\ kg \times 0.6} = 1.09 \frac{MJ_{Ethanol}}{MJ_{fuel}}$$

The ethanol-to-jet plant in Geleynse et al [13] has an ethanol input rate of 200 ton/day, or 181.4 tonne/day. Assuming this plant has a capacity factor of 0.85, the annual mass-based fuel production rate is:

$$Production\ Rate\ (mass) = 181.4 \frac{tonne}{day} \times 365 \frac{day}{yr} \times 0.6 \times 0.85 = 33775 \frac{tonne}{yr}$$

By further assuming the fuel product contains 82% SPK and 18% naphtha, the annual volume-based fuel production rate is:

$$Production\ Rate\ (volume) = 33775 \frac{tonne}{yr} \times (0.82 \times 323.6 \frac{gal}{tonne} + 0.18 \times 352.1 \frac{gal}{tonne}) \times \frac{1\ million\ gal}{10^6\ gal} = 11.1 \frac{million\ gal}{yr}$$

For other SLF production pathways evaluated in this study, their FTS blocks, adopted from Greig et al [14], can produce 3525 bbl SLF/day, which is equivalent to 45.9 million gal SLF/year (capacity factor = 0.85). To maintain consistency in the analysis across our evaluated pathways, we scaled up the ethanol-to-jet plant in Geleynse et al [13] by a factor of 4.14 (45.9/11.1). The CAPEX of the resulting plant is estimated as the product of the CAPEX (excluding working capital) of the original plant (23.3 million 2022$ [13]) and the capacity ratio to power 0.68, as reported by Crawford [15]:

$$CAPEX = 23.3 \times 4.14^{0.68} = 61.2\ million\ 2022\$$$

To estimate OPEX, we considered VOMC and FOMC separately. VOMC of 13.7 million 2022$/y consists of the cost of utilities, catalysts and hydrogen. The cost of the first two items is obtained by linearly scaling (by 4.14 factor) the corresponding values for the original plant, and the cost of hydrogen is determined using the mass-based hydrogen-to-ethanol ratio of 0.595% and the unit hydrogen cost of 0.31 $/kg (as calculated in our paper for green $H_2$ with IRA subsidy). Our FOMC estimate of 6.1 million 2022$/y consists of the cost maintenance, property taxes and insurance, and plant labor. The calculation of FOMC is not detailed by Geleynse et al [13], so we opted to follow Greig et al [14], who assumed annual maintenance cost to be 2% of CAPEX, and property taxes and insurance to be an additional 2% of CAPEX. The number of operators required on a 24/7 equivalent basis is 20 in Greig et al [14] for a plant with additional major equipment blocks (e.g., air separation unit, gasifier, etc.). We assume the ethanol-to-SLF would be co-located with, and share operating personnel with, ethanol production facilities and there require only 5 additional operators.
Finally, the hourly SLF energy output of the ethanol facilities is:

$$181.4 \frac{tonne}{day} / 24 \frac{hours}{day} * 0.6 * 4.14 * 45.5 \frac{GJ}{tonne} = 854\ GJ/hr\ or\ 237,222\ kW$$

Thus the CAPEX is 258 $/kW$_{SLF,HHV}$, the FOMC is 25.8 $/kW$_{SLF,HHV}$ and the VOMC is 2.15 $/GJ$_{SLF,HHV}$ (with 0.85 capacity factor considered in the latter).

(i) Capital and operating costs for $CO_2$ compression and dehydration that are included in this pathway are not included in costs in this table, but are calculated as described in Table S3 (note *h*) and are included in levelized cost of fuel calculations in the paper.

Table S3. Input energy and material prices ($2022)

| Energy/Material | Value |
|---|---|
| Natural gas Price ($/GJ)[a] | 4.3 |
| Biomass Price ($/GJ)[b] | 6.1 |
| Wind or Solar derived electricity price, incorporating 45Y credit ($/MWh)[c] | 21.5 |
| Average electricity price ($/MWh)[d] | 55 |
| Carbon neutral $CO_2$ from DAC, incorporating 45Q credit ($/$CO_2$)[e] | 164 |
| $CO_2$ transportation cost ($/t $CO_2$)[f] | 19.5 |
| $CO_2$ storage cost ($/t $CO_2$)[f] | 8.5 |
| Ethanol price ($/gal)[g] | 1.83 |
| $CO_2$ compression cost for corn-ethanol CCS[h] | 17 |

(a) The natural gas price assumed in our previous study [5] was $3.2/GJ in 2016 $. The CPI index was applied to express this as $3.89/GJ in 2022 $. With assumed upstream methane emissions is 1.7% (0.29 g of methane per MJ of delivered natural gas, the median value from [16]), the emissions penalty under the IRA is $0.44/GJ. Incorporating the methane penalty into the natural gas price gives 3.88 + 0.44 = $4.3/GJ.

(b) The biomass price is adopted from Larson et al [17] (page 181), which shows the majority of prospective, sustainable U.S. biomass supplies available for delivered costs of $100/metric tonne (2016 $) or less. We adjusted this value to 2022$ using the CPI and converted it to an energy basis assuming a higher heating value of 19.8 GJ/metric tonne. The carbon intensity of biomass is assumed to be 87.9 kg/GJ$_{HHV}$.

(c) The wind and solar-derived electricity price before applying any 45Y credit, $35/MWh in 2016 $, is from Cheng et al [5]. This was escalated to 2022 $ using the CPI. With the 45Y credit ($26/MWh), but derated appropriately (derating factor of 0.81 as determined by Eqn 1 in the main paper), the electricity price is $21.5/MWh.

(d) The average electricity price used by Cheng et al. [5] was $45/MWh in 2016 $. The CPI is applied to convert this to 2022 $.

(e) The price of carbon-neutral $CO_2$ from DAC used by Cheng et al. [5] was $230/t $CO_2$ in 2016 $. With a CPI adjustment, this is $280/t $CO_2$ in 2022 $. The 45Q credit ($130/t $CO_2$), derated appropriately (derating factor of 0.89 as determined by Eqn 2) is subtracted from this to give the cost of $CO_2$ from DAC of $164/t$CO_2$.

(f) $CO_2$ transportation and storage costs are given by Larson et al [17] (page 220) as $16/t and $7/t, respectively, in 2020 $. The CPI is applied to express these in 2022 $.

(g) The ethanol price is based on the median of annual average ethanol prices during 2012 to 2022 (CPI adjusted) [18].

(h) The $CO_2$ compressions and dehydration cost at the corn-ethanol plant is calculated to be 17 $/t $CO_2$. The $CO_2$ concentration in the ethanol fermentation exhaust gas is generally above 98%, which is similar to the $CO_2$ concentration in the gas stream leaving the stripper in amine-based capture processes. Thus, we estimated the cost associated with $CO_2$ compression and dehydration based on NETL's case studies [19, 20] of coal and natural gas-to-electricity with CCS, which use amine-based $CO_2$ capture process. In the coal case, the flowrate of product $CO_2$ at ~150 bar is 1.28 million lb/hr, or 581.3 tonne/hr, and the auxiliary electricity load is 44,380 kW. Hence, the unit electricity consumption is 76.3 kWh/tonne $CO_2$. The TPC of the compression and dehydration block is $86.7M, and the scaling factor is 0.61. In the natural gas case, the flowrate of product $CO_2$ at ~150 bar is 0.49 million lb/hr, or 223.9 tonne/hr, and the auxiliary electricity load is 17,090 kW. Hence, the unit electricity consumption is also 76.3 kWh/tonne $CO_2$. The TPC of the compression and dehydration block is $59.7M, and the scaling factor is 0.41. In our scaled-up plant that converts 751 tonne EtOH/day to SLF, the correspoding $CO_2$ flowrate from the fermentation tank is 29.9 tonne/hr, based on the fermentation stoichiometry. Thus, the TPC of the compression and dehydration block at our corn-ethanol plant, estimated from the coal case or the natural gas case above, is $86.7M × (29.9/581.3)$^{0.61}$ = $14.2M or $59.7M × (29.9/223.9)$^{0.41}$ = $26.2 M, respectively. By taking an average of these two values and applying a CRF of 0.131, our estimate of the annualized CAPEX is $2.64 M. The annual $CO_2$ product at the corn-ethanol plant is 0.223 million tonne (capacity factor = 0.85), so the unit CAPEX for $CO_2$ compression and dehydration is $11.9/tonne $CO_2$. As for the OPEX, we assumed that electricity at $60 /MWh is the only input and estimated the OPEX to be $4.6/tonne $CO_2$. So, we used $17/tonne $CO_2$ as the total cost for $CO_2$ compression and storage in relevant calculations.

Table S4. Upstream GHG emissions of feedstocks

| | Unit[a] | 100-year (GWP-100) | REF |
|---|---|---|---|
| Biomass production & harvesting & transportation | kg $CO_2$e/GJ | 5.08 | [21] |
| Natural gas - production, processing, transmission, distribution | kg $CO_2$e/GJ | 13.8 | [16] |
| Corn ethanol | kg $CO_2$e/GJ | 50 | [21] |
| Corn ethanol – with CCS | kg $CO_2$e/GJ | 18 | (b) |

(a) These are per HHV basis

(b) We estimate the life cycle GHG emissions of the ethanol with CCS pathway based on fermentation stoichiometry: 1 mole of glucose ($C_6H_{12}O_6$) forms 2 mole ethanol ($C_2H_5OH$) and 2 mole $CO_2$. So, the maximum amount of $CO_2$ available for capture is equivalent to half of the carbon content in the ethanol. The HHV energy content of ethanol is 29.7 MJ/kg and its carbon content is 1.91 kg $CO_2$e/kg, so the $CO_2$ equivalents in ethanol = 64 kg $CO_2$e/GJ, and the maximum $CO_2$ available for capture = 32 kg $CO_2$/GJ. The fermentation of corn-to-ethanol provides a 99.9 percent pure stream of biogenic $CO_2$ (after dehydration). So, the carbon intensity of the ethanol-CCS pathway = 50 – 32 = 18 kg $CO_2$/GJ.

Table S5 Energy and carbon contents of biomass and fuels [5].

| | Higher Heating Value, $GJ_{HHV}$/tonne | Ratio of Higher to Lower Heating Value | Carbon Content kg $CO_2$/$GJ_{HHV}$ |
|---|---|---|---|
| Biomass (dry) | 19.8 | 1.07 | 87.9 |
| Natural gas | 55.5 | 1.11 | 50 |
| $H_2$ | 142 | 1.18 | 0 |
| SLF | 45.5 | 1.05 | 67.7 |

## 4. Historical jet fuel prices, ethanol prices, RIN values, and LCFS credits

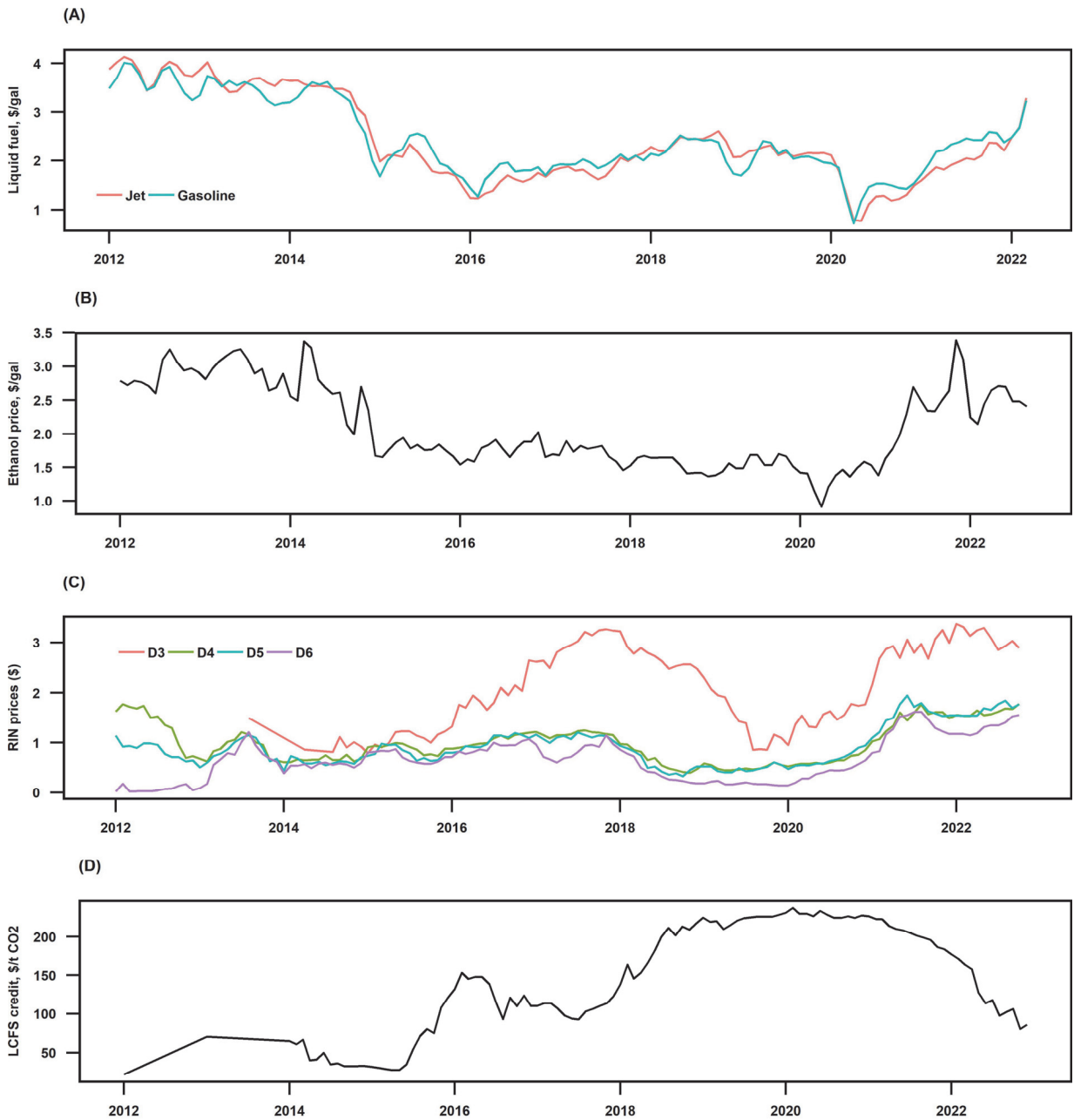

Figure S1. Historical data expressed in constant 2022$ using the CPI. (A) Monthly average refiner resale prices ($/gallon) for kerosene-type jet fuel [22] and gasoline [23]. (B) Monthly average price of Iowa ethanol ($/gal) [18]. (C) Monthly average RIN prices ($/gallon ethanol-energy equivalent) for categories D3 – D6 [24]. (D) Monthly average LCFS credits ($/tCO$_2$ reduction from baseline)[25].

## 5. Unintended consequences of extending 45Z credits?

Here we explore the possibility that 45Z credits might incentivize less-efficient over more-efficient SLF production from biomass at a facility employing CCS.

The IRA stipulates that the 45Z credit is calculated as follows

$$45Z\ credit\ (\$/gal) = \frac{50 - CI_{SLF}}{50} * 1.62\ \$/gal \tag{S3}$$

where $CI_{SLF}$ is the carbon intensity of the fuel pathway (in $kgCO_2e/10^6\ BTU_{SLF}$) and the \$1.62/gallon includes our derating factor (see Method section in main paper). Based on Eqn. S3, the 45Z credit will increase as the CI of the fuel production grows more negative (Figure S2).

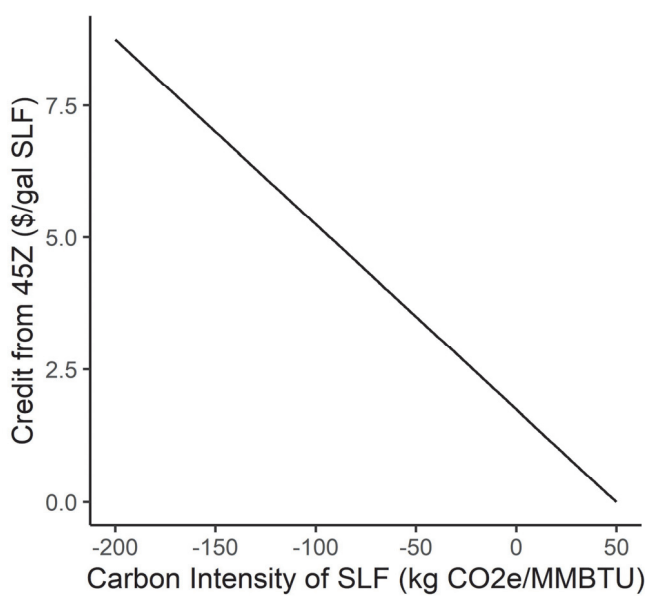

Figure S2. 45Z credit per gallon as a function of SLF lifecycle carbon intensity (calculated from Eqn. S3).

In the absence of any subsidy, a biomass-to-SLF facility with CCS would typically be designed to achieve an energy-conversion efficiency that balances costs (especially capital and feedstock) against expected product revenues. We explore here how the availability of a 45Z credit might impact this calculus. We do this by performing a simplified cost calculation starting with the characteristics of the integrated BioCCS-SLF pathway (P13) in our paper. This plant has a biomass input intensity of 2.0 $GJ_{biomass}$ per $GJ_{SLF}$ (Table S2), corresponding to an energy efficiency of 50% (HHV). We assume a P13-type facility could be designed with a lower efficiency (40%, 30%, or 20%), and that the $CO_2$ capture and storage level would scale inversely with the efficiency. For each level of efficiency, we then estimate the reduction in capital cost that would be needed for that facility to see a net financial benefit from the reduced efficiency, taking into consideration all production costs, anticipated product revenues, and the 45Z credit level. For clarity, we normalize these calculations to one metric tonne of biomass feedstock input.

We assume the one tonne of biomass input has a higher heating value of 18.8 $MMBTU_{HHV}$ and is 47.5% carbon by mass. We again assume the biomass carbon to be carbon-neutral, and for this calculation we neglect the relatively small upstream emissions associated with biomass processing and transport. The carbon intensity (CI) of the SLF product is then calculated by Eqn S4,

$$\mathrm{CI_{SLF}} \left(\frac{kg}{MMBtu}\right) = \frac{\left(-1000 \text{ kg} * 0.475 * \frac{44}{12} + \mathrm{Energy_{SLF}} * 71 \frac{kg\ CO_2}{MMBtu}\right) * 0.87}{\mathrm{Energy_{SLF}}} \quad \text{(S4)}$$

where 71 kg CO$_2$e/MMBTU$_{SLF}$ is the carbon content of the SLF product (see note *e* in Table S2). The factor, 0.87, is the fraction of biomass input carbon available for capture that is captured and stored, which we assume is the same fraction as for our P13 pathway (Table S2), and *Energy$_{SLF}$* is the energy content of the SLF product for each tonne of biomass processed:

$$\mathrm{Energy_{SLF}} \left(\frac{MMBtu_{LHV}}{tonne\ biomass}\right) = 18.8 \frac{MMBtu_{HHV}}{tonne\ biomass} * \eta_{HHV} * \frac{1.05 MMBtu_{LHV}}{MMBtu_{HHV}} \quad \text{(S5)}$$

Where $\eta_{HHV}$ is the assumed biomass energy conversion efficiency for the facility.

The CI value from Eqn S4 is used to calculate the 45Z credits available per one tonne of biomass processed into SLF using Eqn S6.

$$Credit\ 45Z\ (\$) = \mathrm{Energy_{SLF}} \left(\frac{MMBtu_{LHV}}{t\ biomass}\right) * 7.9 \frac{gal}{MMBtu} * \frac{50 - CI_{SLF}}{50} * 1.62\ \$/gal \quad \text{(S6)}$$

Assuming the carbon-negative SLF product has the same market price as its petroleum-derived counterpart (2.2 $/gal for this calculation), the sales revenue for SLF is calculated using Eqn S7.

$$Revenue\ FTL\ (\$) = \mathrm{Energy_{SLF}} \left(\frac{MMBtu_{LHV}}{t\ biomass}\right) * 7.9 \frac{gal}{MMBtu} * 2.2\ \$/gal \quad \text{(S7)}$$

Eqns. S6 and S7 are the revenues that the SLF producer receives.

Next, we calculate production costs. First, we adopt the cost assumptions for input biomass and for CO$_2$ transport and storage directly from Table S3. Next, the fixed costs (CAPEX and FOM) for P13 (Integrated BioCCS-SLF in Table S2), whose energy efficiency η is 50%, are normalized to one tonne of biomass input and levelized (using CRF = 0.131) to give a value of $218/t. The VOM for P13 (Table S2) is converted to $5.43/MMBTU for our calculations. Finally, we assume that fixed costs will decrease by n% for each 10-percentage points reduction in efficiency. With these assumptions, the four levelized-cost components, i.e., biomass input, fixed costs, VOM, and CO$_2$ transport and storage, are calculated by Eqns. S8 through Eqn S11

$$Biomass\ (\$) = \$\ 121 \quad \text{(S8)}$$

$$Fixed_\eta (\$) = \left(\frac{(\eta - 0.5) * n\%}{0.1} + 1\right) * \$\ 218 \quad \text{(S9)}$$

$$VOM\ (\$) = \mathrm{Energy_{FTL}}(MMBtu) * 5.43\ (\$/MMBtu) \quad \text{(S10)}$$

$$CO_2\ T\&S\ (\$) = \left(-1000\ kg * 0.475 * \frac{44}{12} + \mathrm{Energy_{FTL}} * 71 \frac{kg\ CO_2}{MMBtu}\right) * 0.87 * \frac{28\$}{t\ CO_2} * \frac{1t}{1000\ kg} \quad \text{(S11)}$$

These, together with the two revenue components (Eqn S6 and S7), determine the net levelized value per one tonne of biomass, as in Eqn S12.

$$net\ levelized\ value\ \left(\frac{\$}{t\ biomass}\right) = Credit\ 45Z\ (\$) + Revenue\ SLF\ (\$) - Biomass\ (\$) \quad \text{(S12)}$$
$$-Fixed\ (\$) - VOM - CO_2\ T\&S$$

Figure S3 shows calculation results for n% equal to 9.5%. At this level of assumed reduction in fixed costs for each 10 percent reduction in efficiency, the levelized value per tonne of biomass is approximately constant across the assumed full range of efficiencies. On the revenue side of the ledger, as efficiency decreases from 50% to 20%, the CI of SLF goes from -98 to -337 kg $CO_2$e/MMBtu, but the 45Z credit per one tonne of biomass remains almost unchanged because the increase in per-gallon credit that comes with lower efficiencies is offset by reduced gallons of production. The reduced gallons of production also lead to reduced revenue from SLF sales. On the cost side of the ledger, biomass cost is constant since we normalized the calculations to one tonne of biomass. VOM decreases slightly due to reduced output, and $CO_2$ T&S costs increase slightly due to the increased $CO_2$ capture and storage.

From this analysis, we can conclude that the 45Z credit is unlikely to incentivize a less-efficient biomass-SLF with CCS facility over a more-efficient one, as long as the percentage reduction in fixed costs afforded by reducing the design efficiency is not much larger than the number of percentage points by which efficiency is reduced. For example, a facility designed for an efficiency of 20% instead of 50%, would need to achieve fixed cost reductions of at least 30% before seeing a net financial gain.

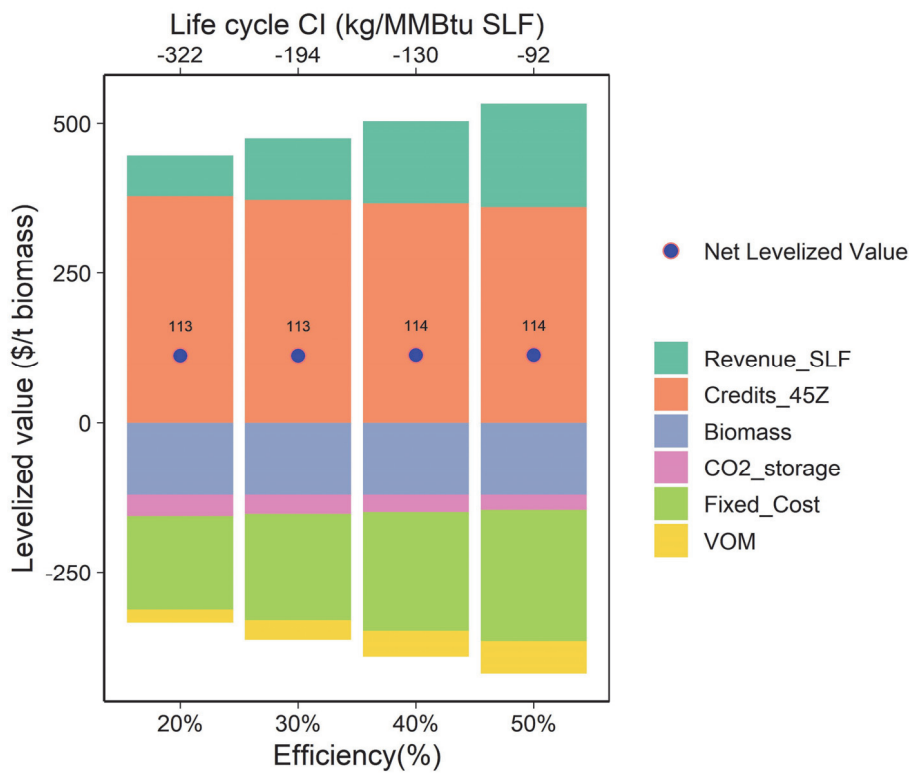

Figure S3. Net levelized value from using one tonne of biomass for SLF production with CCS. The x-axis values are assumed energy efficiencies, and the values above the bars are the corresponding life cycle GHG emissions (kg$CO_2$e/MMBTU$_{SLF}$). When fixed costs are assumed to fall 9.5% for each 10 percent reduction in efficiencies, the levelized value stays roughly constant for different efficiencies in the range shown here.

## 6. Complete design parameter spaces of D5 RIN and D3 RIN.

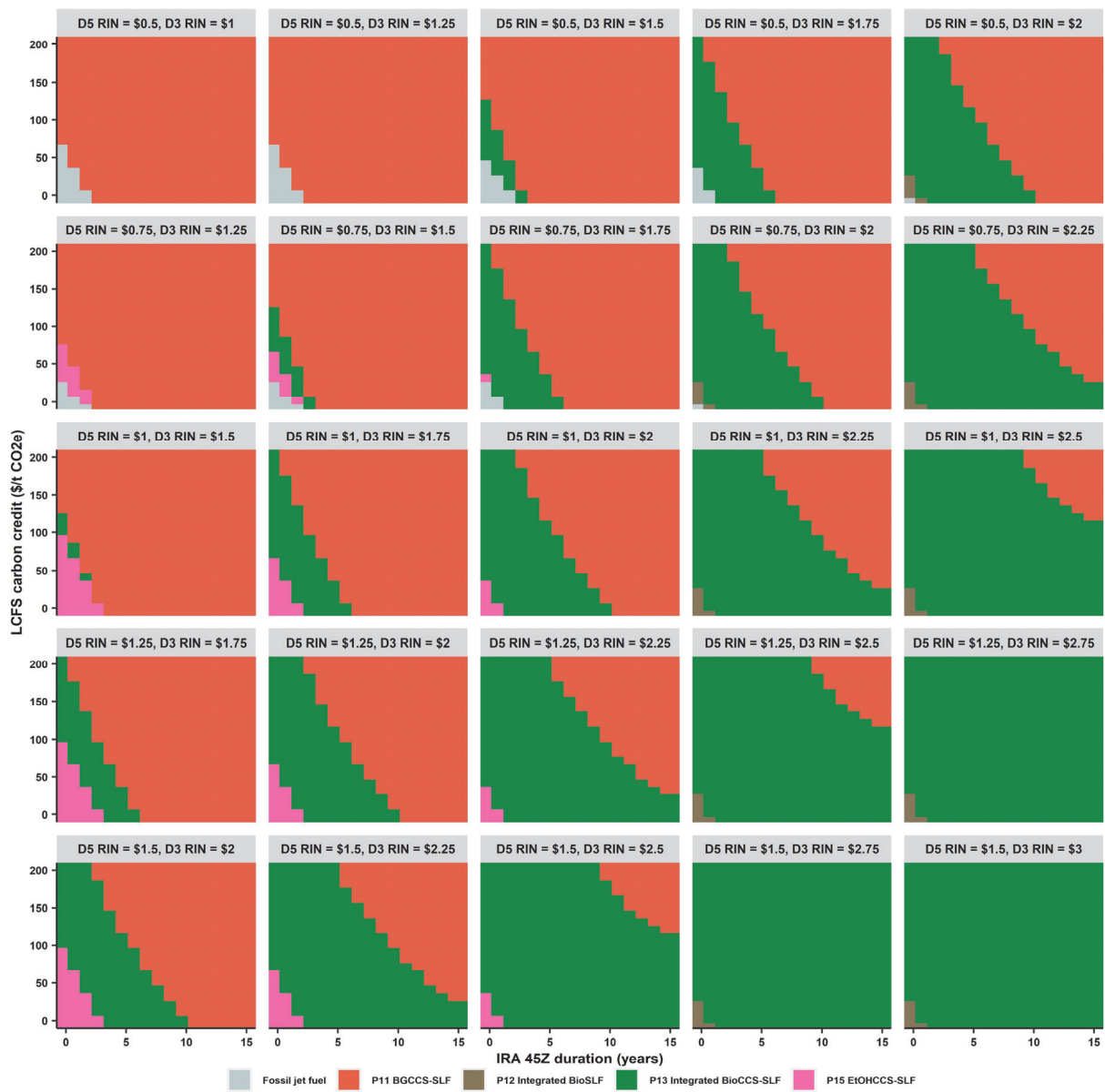

Figure S4. SLF production pathways that would be competitive with fossil jet fuel at 2.2 $/gal under different 45Z durations, LCFS carbon credits, and RIN prices. D5 RIN ranges from $0.5 to $1.5, and D3 RIN ranges from $1 to $3.

## 7. Comparing the value of IRA credits with the Social Cost of Carbon

Figure S5 shows $CO_2$ mitigation subsidies implied by the IRA's 45Q, 45V, and 45Z provisions. Shown for perspective is the range in estimated social cost of carbon for the 2030 to 2040 timeframe [26].

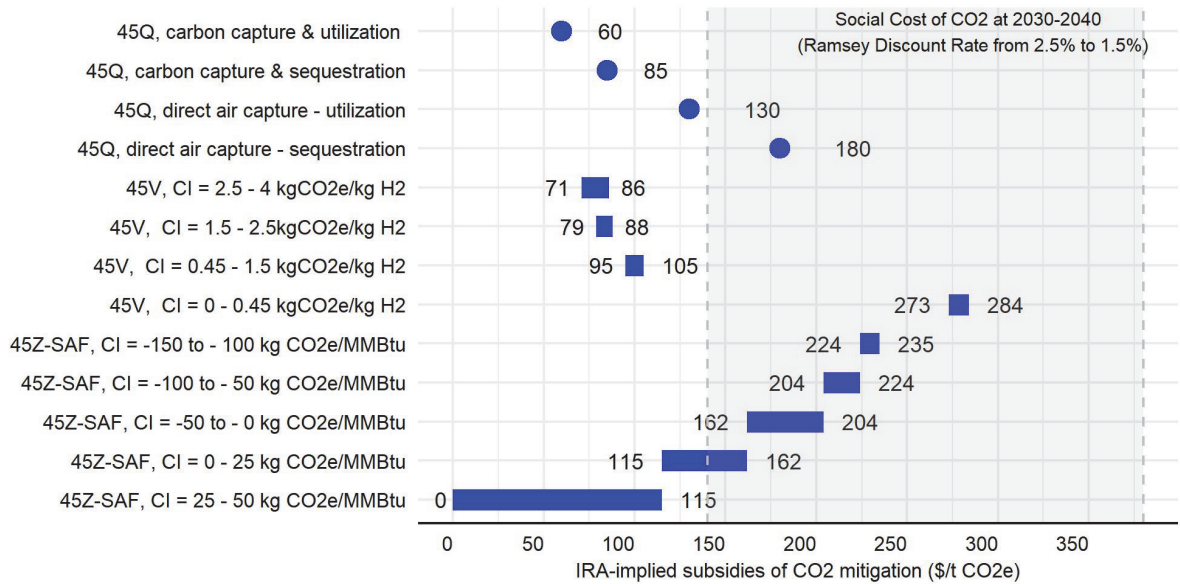

Figure S5. Levelized subsidies for carbon mitigation (LSCM, $/t CO$_2$e), implied in IRA tax credits. Asset life and duration of credits are not accounted here. The CO$_2$ credits for 45V and 45 Z are determined by Eqn 4, wherein the benchmark CI$_{fossil}$ values for hydrogen and SLF are 11 kg CO$_2$/kg (for H$_2$ from SMR) and 85 kg CO$_2$/MMBTU$_{LHV}$ (for conventional jet fuel), respectively. The social cost of carbon range is from [26].

**References:**

1. Jenkins, J.D., Farbes, Jamil, Jones, Ryan, & Mayfield, Erin N., *REPEAT Project Section-by-Section Summary of Energy and Climate Policies in the 117th Congress [Data set]* 2022; Available from: https://doi.org/10.5281/zenodo.6993118.
2. U.S. Environmental Protection Agency, *Renewable Fuel Standard Program*. [cited 2022 Dec 14th]; Available from: https://www.epa.gov/renewable-fuel-standard-program.
3. California Air Resource Board, *Low Carbon Fuel Standard*. [cited 2022 Dec 14th]; Available from: https://ww2.arb.ca.gov/our-work/programs/low-carbon-fuel-standard.
4. California Code of Regulations, *Cal. Code Regs. Tit. 17, § 95484 - Annual Carbon Intensity Benchmarks*. 2022.
5. Cheng, F., et al., *The value of low-and negative-carbon fuels in the transition to net-zero emission economies: Lifecycle greenhouse gas emissions and cost assessments across multiple fuel types.* Applied Energy, 2023. **331**: p. 120388.
6. California Air Resource Board, *LCFS Electricity and Hydrogen Provisions*. [cited 2022 Dec 14th]; Available from: https://ww2.arb.ca.gov/resources/documents/lcfs-electricity-and-hydrogen-provisions.
7. Department of Energy, *DOE Hydrogen and Fuel Cells Program Record*. 2022; Available from: https://www.hydrogen.energy.gov/pdfs/20007-hydrogen-delivery-dispensing-cost.pdf.
8. California Air Resource Board, *Low Carbon Fuel Standard* 2020 [cited 2022 Dec 14th]; Available from: https://ww2.arb.ca.gov/sites/default/files/2020-09/basics-notes.pdf.

9. Kreutz, T.G., et al., *Techno-economic prospects for producing Fischer-Tropsch jet fuel and electricity from lignite and woody biomass with CO2 capture for EOR.* Applied Energy, 2020. **279**: p. 115841.
10. California Code of Regulations, *Cal. Code Regs. tit. 17 § 95482 - Fuels Subject to Regulation.* 2022.
11. U.S. Environmental Protection Agency, *Fuel Pathways under Renewable Fuel Standard*. [cited 2022 Dec 29th]; Available from: https://www.epa.gov/renewable-fuel-standard-program/fuel-pathways-under-renewable-fuel-standard.
12. The Chemical Engineering Plant Cost Index, Chemical Engineering, 2023 [cited 2023 Feb 28th]; Available from: https://www.chemengonline.com/pci-home.
13. Geleynse, S., et al., *The alcohol‐to‐jet conversion pathway for drop‐in biofuels: techno‐economic evaluation.* ChemSusChem, 2018. **11**(21): p. 3728-3741.
14. Greig, C., et al., *Lignite-plus-Biomass to Synthetic Jet Fuel with CO2 Capture and Storage: Design, Cost, and Greenhouse Gas Emissions Analysis for a Near-Term First-of-a-Kind Demonstration Project and Prospective Future Commercial Plants*. United States, 2017, available from https://doi.org/10.2172/1438250.
15. Crawford, J.T., et al., *Hydrocarbon bio-jet fuel from bioconversion of poplar biomass: techno-economic assessment.* Biotechnology for Biofuels, 2016. **9**: p. 1-16.
16. Littlefield, J.A., et al., *Synthesis of recent ground-level methane emission measurements from the U.S. natural gas supply chain.* Journal of Cleaner Production, 2017. **148**: p. 118-126.
17. Larson, E., et al., *Net-Zero America: Potential Pathways, Infrastructure, and Impacts, Final report.* Princeton University, Princeton, NJ, 29 October 2021.
18. Center for Agricultural and Rural Development, *Historical Ethanol Operating Margins*. 2023 [cited 2023 July 6 th]; Available from: https://www.card.iastate.edu/research/biorenewables/tools/hist_eth_gm.aspx.
19. Zoelle, A. and N. Kuehn, *Quality Guidelines for Energy System Studies: Capital Cost Scaling Methodology: Revision 4 Report*. 2019, National Energy Technology Laboratory (NETL), Pittsburgh, PA, Morgantown, WV ….
20. Robert, E., et al., *Cost and performance baseline for fossil energy plants volume 1: bituminous coal and natural gas to electricity*. 2019, National Energy Technology Laboratory (NETL), Pittsburgh, PA, Morgantown, WV ….
21. Wang, M., et al., *Greenhouse gases, Regulated Emissions, and Energy use in Technologies Model ® (2022 .Net)*. 2022: United States.
22. U.S. Energy Information Adminstration, *U.S. Kerosene-Type Jet Fuel Wholesale/Resale Price by Refiners*. [cited 2022 Dec 15th]; Available from: https://www.eia.gov/dnav/pet/hist/LeafHandler.ashx?n=pet&s=ema_epjk_pwg_nus_dpg&f=m.
23. U.S. Energy Information Adminstration, *U.S. Total Gasoline Wholesale/Resale Price by Refiners*. [cited 2022 Dec 15th]; Available from: https://www.eia.gov/dnav/pet/hist/LeafHandler.ashx?n=PET&s=EMA_EPM0_PWG_NUS_DPG&f=M.
24. Environmental Protection Agency, *RIN Trades and Price Information*. [cited 2022 Dec 15th]; Available from: https://www.epa.gov/fuels-registration-reporting-and-compliance-help/rin-trades-and-price-information.
25. California Air Resource Board, *Monthly LCFS Credit Transfer Activity Reports*. [cited 2022 Dec 15th]; Available from: https://ww2.arb.ca.gov/resources/documents/monthly-lcfs-credit-transfer-activity-reports.
26. U.S. Environmental Protection Agency, *Supplementary Material for the Regulatory Impact Analysis for the Supplemental Proposed Rulemaking, "Standards of Performance for New,*

*Reconstructed, and Modified Sources and Emissions Guidelines for Existing Sources: Oil and Natural Gas Sector Climate Review"*. 2022; Available from: https://www.epa.gov/system/files/documents/2022-11/epa_scghg_report_draft_0.pdf.